\documentclass[twocolumn,tighten,times]{aastex62}
\usepackage{mathrsfs}
\usepackage{amsmath}
\usepackage{url}
\usepackage[normalem]{ulem}
\usepackage{graphicx}
\usepackage{float}
\usepackage{comment}
\usepackage{mathtools}
\usepackage{amsmath}

\usepackage{natbib}
\usepackage{color}
\bibpunct{(}{)}{;}{a}{}{,}

\newcommand\aproxgt{\mathrel{%
      \rlap{\raise 0.511ex \hbox{$>$}}{\lower 0.511ex \hbox{$\sim$}}}}
\newcommand\aproxlt{\mathrel{%
      \rlap{\raise 0.511ex \hbox{$<$}}{\lower 0.511ex \hbox{$\sim$}}}}

\newcommand{\ignore}[1]{}
\newcommand{\storm}{{STORM}}
\newcommand{\swift}{{\it Swift}}
\newcommand{\hst}{{\it HST\,}}
\newcommand{\xmm}{{\it XMM-Newton\,}}
\newcommand{\nustar}{{\it NuSTAR\,}}

\newcommand{\et}{{et~al.\, }}

\newcommand{\hb}{\mbox{\rm H$\beta$}}

\newcommand{\civ}{\mbox{\rm C\,{\sc iv}}}

\newcommand{\siIV}{\mbox{\rm Si\,{\sc iv}}}
\newcommand{\heii}{\mbox{\rm He\,{\sc ii}}}

\begin{document}

\title{Space Telescope and Optical Reverberation Mapping Project. XIII. 
An Atlas of UV and X-ray Spectroscopic Signatures of the Disk Wind
in NGC 5548}

\author[0000-0002-0964-7500]{M.~Dehghanian}
\affiliation{\ignore{UKy}Department of Physics and Astronomy, The University of Kentucky, Lexington, KY 40506, USA}

\author[0000-0003-4503-6333]{G.~J.~Ferland}
\affiliation{\ignore{UKy}Department of Physics and Astronomy, The University of Kentucky, Lexington, KY 40506, USA}

\author[0000-0001-6481-5397]{B.~M.~Peterson}
\affiliation{Space Telescope Science Institute, 3700 San Martin Drive, Baltimore, MD 21218, USA}
\affiliation{Department of Astronomy, The Ohio State University, 140 W 18th Ave, Columbus, OH 43210, USA}
\affiliation{Center for Cosmology and AstroParticle Physics, The Ohio State University, 191 West Woodruff Ave, Columbus, OH 43210, USA}

\author[0000-0002-2180-8266]{G.~A.~Kriss}
\affiliation{Space Telescope Science Institute, 3700 San Martin Drive, Baltimore, MD 21218, USA}

\author[0000-0003-0944-2008]{K.~T.~Korista}
\affiliation{\ignore{WM}Department of Physics, Western Michigan University, 1120 Everett Tower, Kalamazoo, MI 49008-5252, USA} 

\author[0000-0002-2908-7360]{M.~R.~Goad}
\affiliation{\ignore{Leicester}Department of Physics and Astronomy, University of Leicester,  University Road, Leicester, LE1 7RH, UK}

\author[0000-0002-8823-0606]{M.~Chatzikos}
\affiliation{\ignore{UKy}Department of Physics and Astronomy, The University of Kentucky, Lexington, KY 40506, USA}

\author[0000-0002-2816-5398]{{M.~C.~Bentz}}
\affiliation{\ignore{Georgia}Department of Physics and Astronomy, Georgia State University, 25 Park Place, Suite 605, Atlanta, GA 30303, USA} 
\author[0000-0002-2915-3612]{F.~Guzm\'{a}n}
\affiliation{ Department of Physics and Astronomy, The University of North Georgia, Dahlonega, GA 30597 USA}

\author[0000-0002-4992-4664]{M.~Mehdipour}
\affiliation{\ignore{SRON}SRON Netherlands Institute for Space Research, Sorbonnelaan 2, 3584, CA Utrecht, The Netherlands}

\author[0000-0003-3242-7052]{G.~De~Rosa}
\affiliation{Space Telescope Science Institute, 3700 San Martin Drive, Baltimore, MD 21218, USA}

\shorttitle{the disk winds}
\shortauthors{Dehghanian \et}

\begin{abstract}
The unusual behavior of the spectral lines of NGC5548 during the STORM campaign demonstrated a
missing piece in the structure of  AGNs. For a two-month
period in the middle of the campaign, the
spectral lines showed a deficit in flux and a reduced response to the variations of the UV continuum. This was the first time
that this behavior was unequivocally observed in an AGN.
Our previous papers
explained this as
being due to a variable disk-wind which
acts as a shield and alters the SED.   
Here we use Cloudy to create an atlas of photoionization
models for a variety of disk-winds to
study their effects on the SED.  We show that the winds have
three different cases:
Case 1
winds are transparent, fully ionized and have minimal effects on the
intrinsic SED, although they can produce some line emission,
especially \heii\  or FeK$\alpha$.  We propose thatthis is the situation in
most of the AGNs. 
Case 2 winds have a He$^{++}$-He$^{+}$ ionization-front, block
part of the XUV continuum but transmit much of the Lyman continuum.  
They lead to the observed abnormal behavior. 
Case 3 winds have H$^{+}$ ionization-front and block much of the 
Lyman continuum.  
The results show
that the presence of the winds has important
effects on the spectral lines of AGNs. They
will thus have an effect on the measurements of the black hole mass and the geometry
of the AGN.  
This atlas of spectral simulations can serve as a guide to future reverberation campaigns.

\end{abstract}


\section{INTRODUCTION } 

Active galactic nuclei (AGNs) provide one of the best
tools to trace the evolution of galaxies and to apply
constraints on the structure of the cosmos. They are
the compact central regions of massive galaxies and the
most luminous objects in the universe. Because of their
high luminosity, they are detectable at very high
redshift. Their brightness results from the accretion
of matter into a supermassive black hole (SMBH) at
their center. A long-standing goal of AGN research has
been to understand the mass and inner structure through gas flows in AGNs; clearly, accreted gas powers
the AGN itself and outflows must interact with the
surrounding galaxy, and the details of these
interactions have implications for galaxy evolution.
Unfortunately, the gas flows within the black hole
radius of influence are largely unresolved, which complicates
our attempt to understand their structure and
interactions.

Reverberation mapping (RM) is the fundamental method
for determining the inner structure and mass of AGNs by
use of temporal variations as a tool to study the spatially unresolvable flows and structures. The continuum
emission that originates in the accretion disk
surrounding the black hole undergoes irregular flux
variations. The broad emission-line fluxes change in
response to these variations, but with a time delay due
to the light-travel time between the accretion disk and
the broad line region (BLR); measurement of these time
delays is the fundamental goal that underlies the
technique of reverberation mapping \citep{BMcK82, Pet93}. Similarly, changes in the intrinsic absorption
features allow us to make inferences about changes in
the AGN spectral energy distribution (SED) as well as
the ionization state, temperature, and density of the
absorbing gas, and other characteristics.

The Seyfert 1 galaxy NGC 5548 has been the target of
many reverberation campaigns
(see \cite{Pet02,DeRosa15}, and references therein).
 In
2013, the “Anatomy” campaign \citep{Kaastra14,
Mehd15, Mehd16, Arav15, Ursini15, Gesu15, Whewell15,
Ebrero16, Cappi16} monitored this
object mainly using \xmm\  and the Neil Gehrels
\swift\ Observatory, enhanced with data from the \hst\  Cosmic Origins Spectrograph (COS). One year later, the Space Telescope and Optical
Reverberation Mapping program, or AGN \storm\ , began
observing the object using Hubble Space Telescope , the
Neil Gehrels \swift\  Observatory, and the Chandra X-Ray
Observatory \citep{DeRosa15, Edelson15,
Fau16, Goad16, Mathur17, Pei17, Starkey17, Deh19a, Kriss19, Horn20,
Deh20}. 

The results from both campaigns found that NGC 5548 was
in an unusual state: In 2013, the Anatomy campaign
revealed that strong and persistent soft X-ray
absorption was present.  This was produced by an
ionized outflow which we term the line-of-sight (LOS)
obscurer.  The obscurer is an outflowing stream of
ionized gas with embedded colder, denser parts \citep{Kaastra14}, blocking a considerable amount of
the soft X-ray emission and causing simultaneous deep,
broad UV absorption troughs \citep{Mehd15}. 

High cadence \hst\  spectroscopy performed by the AGN STORM project in 2014
found that the emission lines strongly decorrelated  from the continuum for 70 of the 180 days of the campaign.  In
the words of the investigators, the emission lines ``went
on holiday''. This kind of decorrelation had never been
commented on before. \cite{DeRosa15} present
details about the observations, and \cite{Goad16}
and \cite{Pei17} give quantitative measurements
of the holiday observed in the emission lines (the
emission-lines holiday). 

Subsequent work found that high-ionization absorption
lines had a similar holiday \citep{Kriss19}.
\cite{Deh19a}, hereafter D19a, showed that
this was due to changes in the covering factor (CF) of
the LOS obscurer, as verified by the \swift\ 
observations.  \cite{Deh19b}, hereafter
D19b, proposed that the LOS obscurer originates in a
disk wind so that it must extend down to the equator.
Ionizing radiation produced by the accretion disk must
pass through the base of this wind, which we term the
equatorial obscurer, before striking the BLR.  D19b
showed that the BLR holiday can be
produced by changes in the density or column density of
the equatorial obscurer. Later, \cite{Deh20}, hereafter D20, used the \hst, \xmm, and \nustar
observations to propose a proper model for the
equatorial obscurer, leading to a model for the disk
wind itself. 

Such disk winds are common although few studies of
their emission and transmission properties have been
made. A number of studies, including \cite{Murray95, le04,Shemmer15, Revalski18}, have invoked translucent screens
that block part of the ionizing radiation to explain
different parts of the AGN phenomenon.  This paper is a
systematic study of the transmission and emission
properties of such translucent screens.  

In the following Section, we explore the disk wind and
its properties. 
We begin with
the LOS obscurer in NGC 5548 because this is well
studied and modelled.  
In Section three, we investigate how
the existence of winds with different parameters will
affect the SED transmitted through the wind or the emission originating from the wind.
We identify three distinct scenarios which depend on whether
H, He, or He$^{+}$ ionization fronts are present in the cloud.
Section four is dedicated to setting limits to the
global covering factor of the upper part of the disk
wind and its resulting emission. It is followed by a
discussion of the base of the wind and the emission
lines produced by it.

\section{THE DISK WIND}
While this paper is motivated by the holiday observed
in NGC 5548, we are not trying to model or analyze any
specific observation. We examine several ways by which
these kinds of winds could affect the SED emitted by
the sources and cause a holiday. These results should
apply to the family of AGNs. Our goal is to show that
cloud shadowing can have a dramatic effect on the
spectra, and it must be considered in all AGN studies. 
\subsection{The Obscurers}
Figure 1 of D19b shows the geometry of NGC 5548 and
includes the disk wind. We refer to the upper part of
the wind as the LOS obscurer since it blocks much of
the soft X-rays. The LOS obscurer can be directly
observed and has a well-determined column density, soft
X-ray absorption, and variable covering factor.  This
obscurer affects the absorption lines (D19a), however,
it does not directly affect the emission lines. The LOS
obscurer first appeared in 2011 \citep{Kaastra14} and
began to cover the central source.  The portion of the
source that is covered by the LOS obscurer varies with
time \citep{Mehd16}. During the holiday, it covered $\sim 86\%$ of the X-ray source and $\sim 30\%$ of the UV source \citep{Kaastra14}. 

The base of the wind, launched from the disk, is called the
equatorial obscurer. This obscurer is the one which
affects the BLR emission lines and produces the
emission line holiday. There are no measurements of its
column density or any other physical properties, but it
is likely denser than the LOS obscurer and has a higher
column density since it is closer to the accretion
disk, where it was originally produced. D20 proposed
physical properties for the translucent equatorial
obscurer for which the wind has a constant optical
depth to be located at the minimum threshold to produce
a holiday (based on figure 4 of D19b).  However, a wind
from anywhere else within the Case 2 area of D19b
figure 4, will still produce the holiday, but will have
different properties. 

The SED striking the BLR first passes through the
equatorial obscurer and D19b argues that this filtering
causes the emission-line holiday. This obscurer absorbs
a great deal of the original SED, so its emission may
be significant. Below we discuss this in more detail. 

The hydrogen density and the ionization parameter of
both obscurers are unknown. Regarding the LOS
obscurer, \cite{Kaastra14} derived an ionization parameter of $\log \xi=-1.2\  \rm erg\  cm\  s^{-1}$, however, later \cite{Cappi16}, found a much higher ionization parameter, $\log \xi=0.4-0.8\  \rm erg\  cm\  s^{-1}$.  Recently \cite{Kriss19} suggested a
still higher ionization parameter 
of $\log \xi=0.8-0.95\  \rm erg\  cm\  s^{-1}$.  

We note that there are two different ionization
parameters: $\xi$ and U, which have been used in various
STORM papers.
The ionization parameter $\xi$ is defined as \citep{Kallman01}:

\begin{eqnarray}
\xi & = & (4 \pi)^2 \int_{1Ryd}^{1000Ryd} \frac{J_\nu\,d\nu}{n({\rm H})} =
\frac{F_{\rm ion}}{n({\rm H})} \nonumber \\
& = & \frac{L_{\rm ion}}{n({\rm H})r^{2}} \ {\left[{\rm erg\,cm\,s}^{-1}\right]}
\end{eqnarray}

while the ionization parameter U is dimensionless and is defined by:

\begin{equation}
\rm U =\frac{Q({\rm H})}{4 \pi r^2 n(H) c},
\end{equation}
in which Q(H)=$\int_{1Ryd}^{\infty} \frac{L_\nu\,d\nu}{h\nu} $ is the total ionizing
photon luminosity.
For the unobscured SED of NGC 5548,
$\log\ \rm U = \log \xi - 1.6$.

As figure 3 of D19b shows, the SED can be dramatically affected when the hydrogen density gets more substantial. The situation may be the same if other parameters, like metallicity, change.  This paper examines a range of obscurer parameters to check for observed properties and possible predictions. 

\subsection{The Covering Factors and Implications for Explaining the Holiday }

Three different types of covering factors will enter in
the following discussion. First, the line of sight
covering factor, LOS CF.  It is the fraction of the
continuum source covered, as seen from our line of
sight.  This covering factor is directly measured
through the hardness ratio estimations based on \swift\
observations \citep{Mehd16}. Second, the global
covering factor, GCF, the fraction of the sky covered
by a cloud, as seen from the central object \citep{Wang12}. This type of CF is not in our LOS, and it
matters when we study the total emission luminosity of
an obscurer emission-line holiday.  Third, the ensemble covering factor, ECF.
This covering factor accounts for the total portion of
the source covered by all clouds in all directions. The
ensemble global covering factor is typically 20$\%$ and
can be determined from the equivalent width (EW) of
emission lines \citep{Osterbrock06}.

D19a describes a physical model that explains the
ab\-sorp\-tion-line holiday as a result of changes in the
LOS CF of the obscurer. In this model, the SED passes
through the LOS obscurer before striking the absorption
cloud. There is a minimal transition in EUV \footnote{We refer to the region 6 -- 13.6 eV (912\,\AA\ to 2000\,\AA)
as FUV; 13.6 -- 54.4 eV (228\,\AA\ to 912\,\AA) as EUV; and 54.4 eV
to few hundred eV (less than 228\,\AA) as XUV.} and soft
X-ray. However, the FUV  part of the SED is almost not
touched. High ionization species are affected by these
changes in a way that the absorption-line holiday
occurs. This explanation is only reliable for the case
of a LOS obscurer and is consistent with 2013 \swift\  observations of NGC 5548. 

The equatorial obscurer shields the BLR, altering the
SED striking it. The ECF of the BLR is unusually large,
~50$\%$, in NGC 5548 \citep[integrated cloud covering fraction;
][]{Korista00}, so the global covering factor of the
equatorial obscurer must also be this large to explain
the BLR holiday (D19b). For this reason, a CF-based
model cannot explain the emission-line holiday. The equatoria
obscurer, which is much closer to the BH than the BLR
and is likely to be the base of the wind, has
instabilities in its mass-loss rate leading to
variations in its hydrogen density. D19b explains that
the variations of the hydrogen density of the
equatorial obscurer can give rise to the emission-line
holiday.

These different covering factors matter because they
define the portion of the spectrum absorbed by the
obscurer. This affects the ability to block the SED
striking the outer clouds, and also determines the emission
from the obscurer. Below we show how changes in the
parameters of the obscurer affect the SED transmitted
through it or emitted by it. 

\section{THE EFFECTS OF DIFFERENT PARAMETERS-GENERAL CONSIDERATIONS}

As mentioned earlier, two obscurers affect the spectrum
of NGC 5548: the LOS obscurer with its impact on
intervening absorption lines, and the equatorial
obscurer with its impact on the broad emission line
clouds.  These obscurers have many free parameters and
variations of any of these parameters will change the
transmitted SED. This section considers the effects of
these parameters starting from a standard model of the
LOS obscurer.  This offers a good starting point since
it is the one with direct measurements and modeling of
its absorption properties. Surprisingly, there 
have been very few systematic explorations of the physical
properties of absorbing gas and the effects of such absorbers on the SED
transmitted through them (e.g. see \cite{Ferland82, Kraem99, le04}.

\subsection{A Typical Cloud }

\cite{Mehd15} derived a standard or baseline model for
the LOS obscurer in NGC 5548. Their model suggested that
the obscurer blocked all of the SED between the FUV
(13.6 eV) and the X-ray (1 keV). The complicated
changes in the narrow absorption lines, where their
degree of line-continuum correlation depends on
ionization potential \citep{Kriss19}, suggests that a portion of the
SED may be transmitted through the obscurer. The EUV and XUV portions of the
transmitted SED are incident upon and  ionize the UV absorbing cloud. This transmitted SED depends on the obscurer's energy-dependent optical depth, which in turn depends on the the obscurer's metallicity (Z), hydrogen number density n(H), and ionization parameter ($\xi$ or U).

Some of the physical properties of the LOS obscurer are
known since we observe it in absorption, while no
properties of the equatorial obscurer are
observationally identified, although it is arguably
more important since it changes the relative strength and response amplitude of the emission lines. 
D20 used a new approach to propose some parameters for
the equatorial obscurer.  The SED transmitted through or emitted by each of
the obscurers is dependent on their column density, hydrogen density, ionization parameter (or distance from the source), and metallicity. Here we investigate this dependency by
modeling different obscurers with different values for these parameters. 
We start with a typical cloud, assuming a density that
is typical of the BLR, $n(\rm H)=10^{10} cm^{-3}$, and
we adopt the intrinsic SED described in figure 3 of
D19a. We assume solar abundances (photospheric), which
is {\tt cloudy}'s default value \citep{Ferland17}
unless otherwise specified.  We start by keeping the
optical depth at 1 keV constant at the value that was
directly observed. This largely reproduces the obscured
SED shown in \cite{Mehd15}. Please note that
the best related models  assume there are in fact two
separate LOS obscurers with different physical
properties such as column density, ionization
parameter, and covering factor. 

\subsection{Varying the Ionization Parameter}

We predict transmitted and emitted SEDs by changing the
ionization parameter ($\xi$) but keeping all other
quantities constant. 
Changes in the ionization parameter are equivalent to changes in the luminosity of the source. Thus, this section  also accounts for variations of the central source luminosity.

\begin{figure*}
\centering
\includegraphics [width=\textwidth]{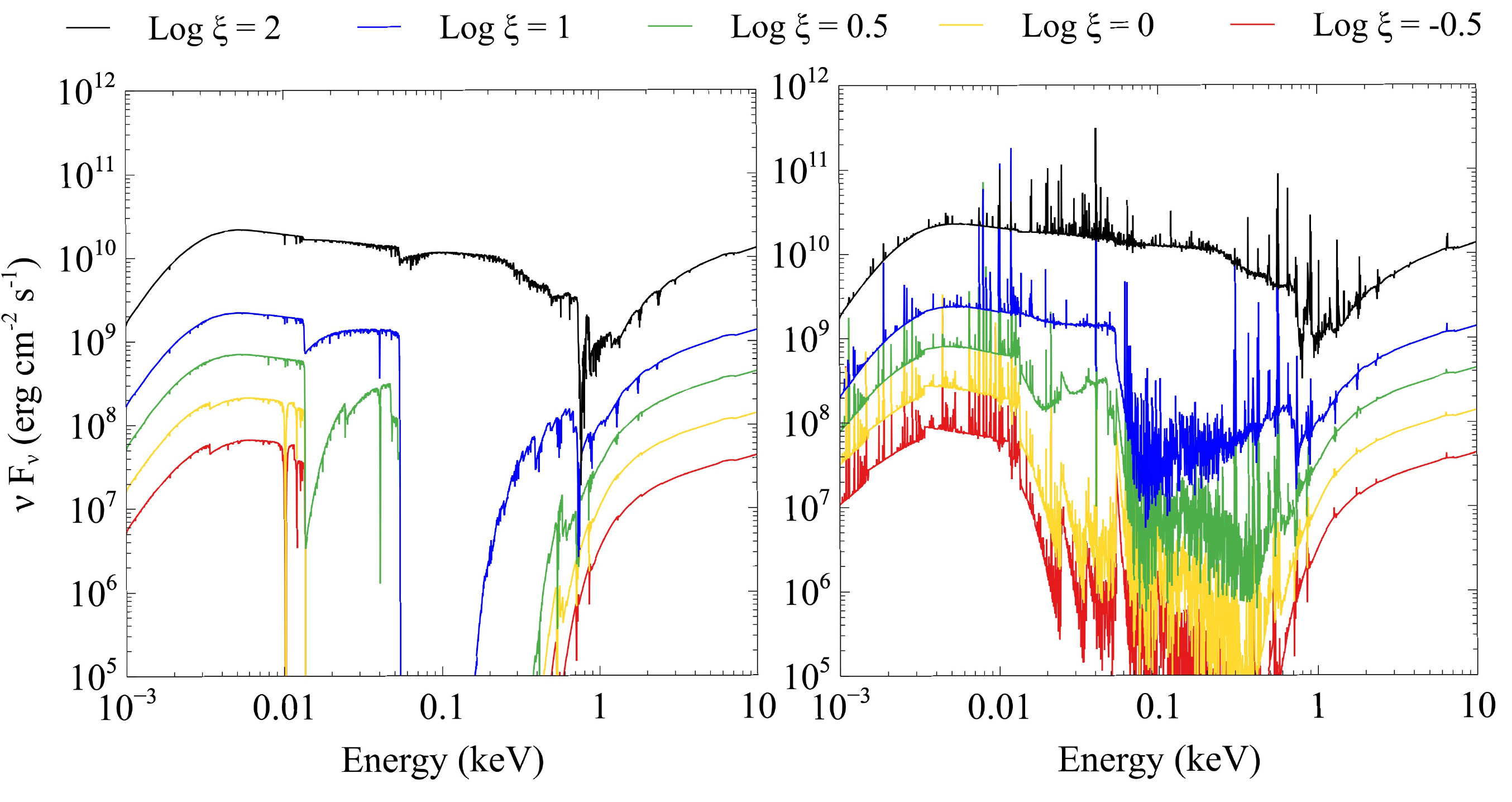}
 \caption{Left panel: The obscurer’s transmitted SED. As the lines show the transmitted SED
 is very sensitive to the ionization parameter in soft X-ray energies. Right panel: total emission(sum of the transmitted and reflected continua plus attenuated incident continuum) from the
 obscurer for different values of the ionization parameter. The
attenuated incident continuum is included. The spectra represent $\log \xi =$2, 1, 0.5, 0,
and $-0.5$ from top to bottom, and
their total hydrogen column densities are 
$\sim 10^{22.7},10^{22.3},10^{22.1},10^{22}, \rm and\  10^{21.4} \rm
cm^{-2}$, respectively. To create
these models, the optical depth at 1 keV is
kept constant and equal to the observed
value.  There is an
animation associated with this Figure which
dynamically illustrates the left panel of the
Figure above. The animation shows how the transmitted SED responses to the variations of the ionization parameter for a range of $-1.5\   \rm erg\  \rm cm\  \rm s^{-1} \leq \xi  \leq +2\  \rm erg\  \rm cm\  \rm s^{-1}$.The real time duration of the video is 6 seconds.} \label{fxi}
\end{figure*}

To keep the thickness of the cloud
constant, we kept the optical depth constant.
For all of the models,
an outer cloud boundary was set so that the absorption optical
depth at 1 keV is $\tau_{abs}\approx$1.9, consistent with
that observed (figure 7 of \cite{Mehd15}). These
models do not necessarily illustrate the obscurers of
NGC 5548 but show how sensitive the transmitted/emitted
SEDs could be to the variations of the ionization
parameter. We ran {\tt cloudy} \citep{Ferland17} for a
grid of ionization parameters between -1.5 and 2 with
steps as small as 0.025 dex. Results can be seen
dynamically in an animation available as a supplement to this paper. Figure~\ref{fxi} shows
examples of both the transmitted SED and the total
emission from the obscurer for five different values of
the ionization parameter $\xi$. 
Please note that in this paper, all the Figures which
are labeled to be the emission from the obscurer are
actually the total emission, which is the sum of the
transmitted and reﬂected continua and lines and assume
100$\%$ covering factor. The attenuated incident
continuum is also shown.

As the left panel of Figure~\ref{fxi} shows, the
transmitted SED is spectacularly dependent on the
value of the ionization parameter:  clearly, the
EUV part of the SED strongly depends on the
ionization parameter while the XUV is always
strongly attenuated unless we adopt a very high
ionization parameter.  For some ionization
parameters, the EUV region is totally blocked, and
for others, transmitted. When it comes to the
emission from the obscurer, this dependency seems
not to be as much as the transmitted SED, however,
the higher the $\xi$ is the more emission is predicted.

Figure~\ref{fxi}, left panel, shows three different styles
of transmitted SED.  These correspond to three
different ionization states for hydrogen and
helium within the obscurer. In order of decreasing
ionization parameter, which lowers the ionization,
increases the opacity, and decreases the
transmission, these are:

\begin{itemize}
  \item Case 1) Very high $\xi$ and ionization: 
  Hydrogen is highly ionized, and the SED is fully
  transmitted.  There is no ionization front in
  this case. The higher ionization means that all
  of the EUV and XUV passes through the cloud.
  There is not enough singly ionized helium to
  fully absorb the XUV. In other words, the
  obscurer is transparent.
  \item Case 2) Intermediate to high $\xi$ and
  ionization:  Some of the EUV is transmitted,
  while little XUV light is transmitted. In this
  case, there is no He$^{0}$-He$^{+}$ionization front, but there is a He$^{+}$ - He$^{++}$
  ionization front.  Enough atomic hydrogen is
  present to block much of the EUV, and the XUV is
  fully blocked by large amounts of singly ionized
  helium. In some samples of Case 2 (green line),
  there is a shallow H$^{0}$ – H$^{+}$ ionization
  front, however in other samples (blue), there is
  almost no hydrogen ionization front. 
  \item Case3) Low $\xi$, low ionization: All EUV and XUV
  light is blocked.  There is an H$^{0}$ ionization
  front.  There are significant amounts of atomic
  H that absorb much of the ionizing continuum.  This is similar to the \cite{Mehd15}
  standard model of the obscurer. This is also the
  case for the LOS obscurer during the AGN STORM
  campaign and was assumed by \cite{Arav15} and
  D19a. 
\end{itemize}

All the SEDs resulting from the full range of
the adopted ionization parameters fall into one
of these categories. Below we show that it does not
matter which parameter of the obscurer is
changing; the resulting transmitted SED will
always be one of these three cases. 

It is worth
emphasizing that although we are motivated by the obscurers in NGC 5548 and we use this AGN as our point of reference, our discussion provides a framework for future AGN modeling. 
We study the effects of different obscurer properties on the transmitted SED.
While all of the presented simulations use the properties of NGC 5548 and its obscurers as input, 
the approach should have a broader application. 

\subsubsection{A physical interpretation of three transmitted cases}
\begin{figure*}
\centering
\includegraphics [width=\textwidth]{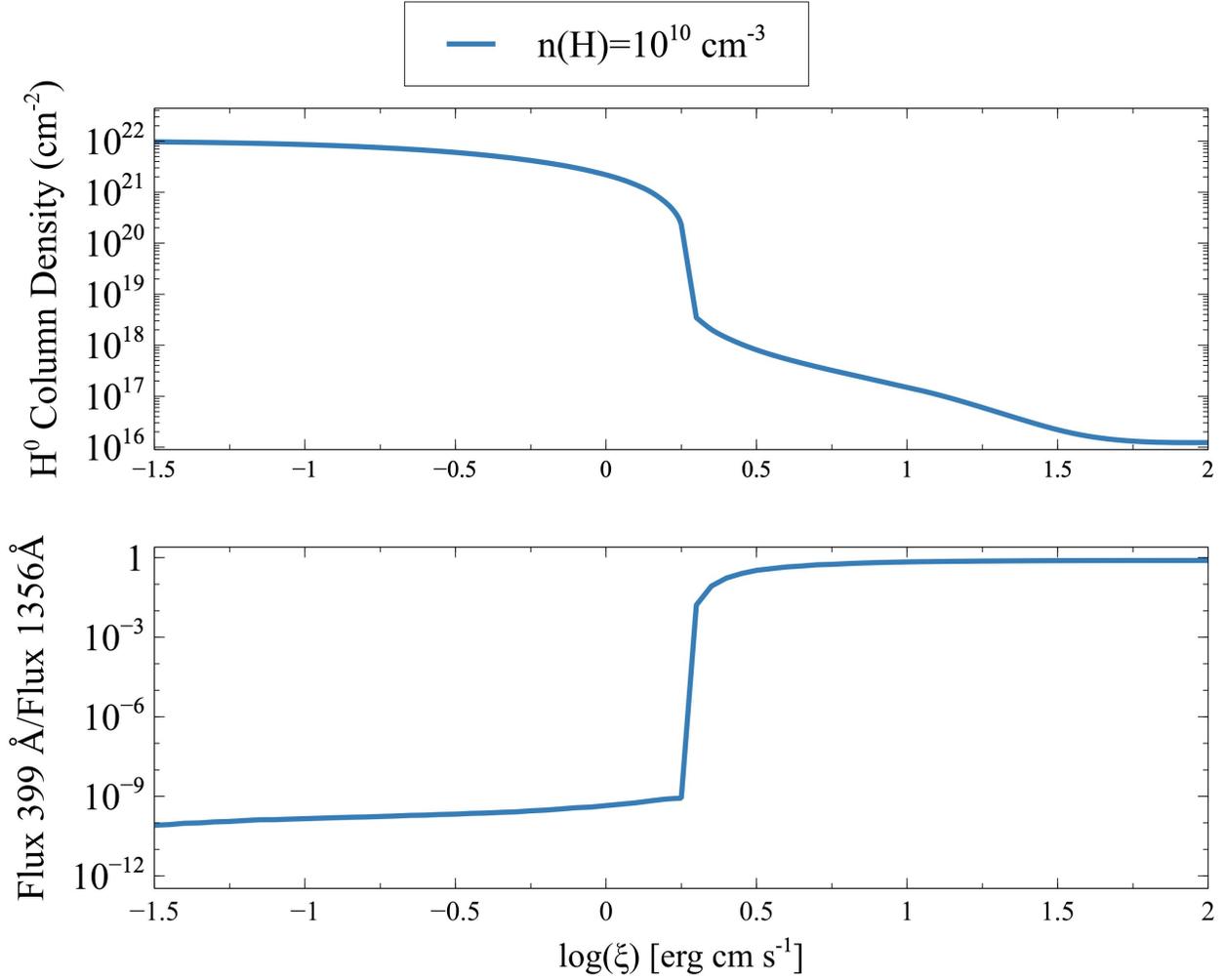}
 \caption{Top panel shows the variation of atomic hydrogen column density as the ionization parameter changes. The bottom panel shows the ratio of 399\AA/1365\AA \ as a function of the ionization parameter.}\label{fxiRatio}
\end{figure*}

We quantify the effects of these changes on the
SED in Figure~\ref{fxiRatio}, which investigates variations of
the ionization parameter, the independent axis,
while the optical depth is kept constant.  The
upper panel gives the H$^{0}$ column density.  The
lower panel shows the ratio of the intensity of
the transmitted continuum at 399~\AA\  (EUV) relative to that at 1356~\AA\  (FUV).  We selected
these two wavelengths because they belong to two
separate energy regions with very different
responses to the changes in the ionization
parameter, so they quantify the changes between
the Cases described above.  \hst\  measures the 1356~\AA\  point while we expect that a photoionized
cloud is most affected by the 399~\AA\  point. This
shows that the 399~\AA\  continuum is strongly
extinguished for low ionization parameters,
corresponding to Case 3.  The abrupt change in the
399~\AA\  transmission and the atomic hydrogen
column density occurs at the ionization parameter
where there is no longer a hydrogen ionization
front, and the cloud is highly ionized.  This is
the transition from Case 3 to Case 2.  If the
extinction changes due to a transition between
Case 3 and 2, then the ionization state of the
absorption cloud will not directly track the \hst\
continuum.

\subsubsection{An upper limit for the ionization parameter}

It is possible to calculate the upper limit of
ionization parameter U (and so $\xi$) for which
each of these cases could be possible. This will
help to adjust the limit on the ionization
parameter based on the model.  Below we calculate
three different maximum value ionization
parameters for which a cloud could have ionization
fronts and so modify the transmitted SED. We
assume that the cloud has a column density of
$\rm N(\rm H)=1.2\times 10^{22} \rm cm^{-2}$. We also
assume that the nebula is optically thick with
a temperature of $\rm T=10^{4} K$. 

The transmitted continuum has to be opaque at the
hydrogen edge to have an H$^{0}$-H$^{+}$ ionization front.
This means that all the photons with energies more
than 1 Rydberg are absorbed by the neutral
hydrogen and an electron and H$^{+}$ are produced
(H$^{0}$+$\gamma$ $\longrightarrow$H$^{+}$+e $^{-}$) also there is an equilibrium
balance. Photoionization balance is the detailed
balance between photoionization and recombination
by electrons and ions \citep{Osterbrock06}:

\begin{equation}
\begin{aligned}
\phi(\rm H) =n_{e} \alpha_{B}(H,T)n_{H^{+}} L_{Stromgren} \ \rm (photons \rm\  s^{-1} \rm cm^{-2})\\ 
=\rm n_{e} \alpha_{B}(H,T) N_{IF}(H^{+}) \ \rm (photons \rm\  s^{-1} \rm cm^{-2}),
\end{aligned}
\end{equation}
in which IF stands for the ionization front and L$_{\rm Stromgren}$ 
is the physical thickness of the Stromgren sphere (H II \ region).

\begin{equation}
\Rightarrow \rm N_{IF}=\frac{\phi(\rm H)}{n_{e} \alpha_{B}(H,T)}
\end{equation}

Since it is always the case that N$_{\rm IF}<$N:
\begin{equation}
\frac{\phi(\rm H)}{\rm n_{e} \alpha_{B}(\rm H,T)}<1.2\times 10^{22} \rm cm^{-2}
\end{equation}
 and:
 
 \begin{equation}
\phi(\rm H)=\rm U\  c\  n_{e} 
\end{equation}

\begin{equation}
\Rightarrow  \frac{\rm U c }{\alpha_{B}(\rm H,T)  }<1.2\times 10^{22} \rm cm^{-2}
\end{equation}

\rm while, 
$\alpha_{B}(\rm H,10^{4} \hbox{K})=2.59 \times 10^{-13}  \rm cm^{3} s^{-1}$, \rm 

so \hbox{for the assumed column density} :

\begin{equation}
\Rightarrow \rm U<10^{-0.98} \rm 
\end{equation}

This is the upper limit in the hydrogen ionization parameter that ensures a hydrogen ionization front. We can always use the
relationships explained in D19a to transform
between U and $\xi$ (for the SED of NGC 5548: $\log \rm U \approx \log \xi-1.6$ ) .

The exact same discussion works for the He$^{0}$-He$^{+}$ ionization front, in which the equilibrium state is:

\begin{equation}
\begin{aligned}
\phi(\rm He^{0})=n_{e} \alpha_{B}(He^{0},T)n_{He^{+}} L_{Stromgren}
\\=\rm n_{e} \alpha_{B}(He^{0},T) N_{IF}(He^{+})
\end{aligned}
\end{equation}
We assume the cosmic abundance ratio of He/H to be
almost 10\%. This results in a helium column density
of $\rm N(He^{+}) =0.1\times N(\rm H)$, so to have a helium ionization front:

\begin{equation}
\phi(\rm He^{0})=n_{e} \alpha_{B}(He^{0},T) N_{IF}(H)\times 0.1
\end{equation}

\begin{equation}
\begin{aligned}
\frac{\phi(\rm H)}{\rm n_{e}}\times \frac{\phi(\rm He^{0})}{\phi(\rm H)}\times \frac{1}{\alpha_{B}(\rm He^{0},T)\times 0.1}=\\
\rm U\  c \times \frac{\phi(\rm He^{0})}{\phi(\rm H)}\times \frac{1}{\alpha_{B}(He^{0},T)\times 0.1}<1.2 \times 10^{22},
\end{aligned}
\end{equation}
in which $\alpha_{B}(\rm He^{0},10^{4} \hbox{K})=2.72 \times 10^{-13} \hbox{cm}^3\hbox{s}^{-1}$.

For the adopted SED and an obscurer located at r $\approx10^{18}$ cm (to be consistent with the LOS obscurer), we find: 
\begin{equation*}
\begin{aligned}
\phi(\rm H)=2.27\times 10^{17} \rm cm^{-2}s^{-1}\\
\phi(\rm He^{0})=1.12\times 10^{17} \rm cm^{-2}s^{-1}\\ 
\phi(\rm He^{+})=4.22\times 10^{16} \rm cm^{-2}s^{-1}\\
\end{aligned}
\end{equation*}

These result in:
\begin{equation}
\Rightarrow  \rm U<10^{-1.65} \rm 
\end{equation}

And similarly, to have a He$^{+}$-He$^{++}$ ionization front:

\begin{equation}
\begin{aligned}
\phi(\rm He^{+})=n_{e} \alpha_{B}(\rm He^{+},T) N_{IF}(He^{++})
\end{aligned}
\end{equation}
for which $\rm N_{IF}(He^{++})\approx N_{IF}(He^{+}) =0.1\times N(H)$ and $\alpha_{B}(\rm He^{+},10^{4} \hbox{K})=1.5 \times 10^{-12} \hbox{cm}^3\hbox{s}^{-1}$, so:

\begin{equation}
\begin{aligned}
\frac{\phi(\rm H)}{n_{e}}\times \frac{\phi(\rm He^{+})}{\phi(\rm H)}\times \frac{1}{\alpha_{B}(\rm He^{+},T)\times 0.1}=\\
\rm U\  c \times \frac{\phi(\rm He^{+})}{\phi(\rm H)}\times \frac{1}{\alpha_{B}(He^{+},T)\times 0.1}<1.2 \times 10^{22}
\end{aligned}
\end{equation}

\begin{equation}
\Rightarrow  \rm U<10^{-0.48} \rm .
\end{equation}

Table~\ref{T1} summarizes the results. Please note that if we use a different location for the obscuring cloud, all the values for $\phi(\rm H)$, $\phi(\rm He^{0})$, and $\phi(\rm He^{+})$ will change accordingly and by the same scale. Thus, based on equations 11 and 14, the final results will stay unchanged.

\begin{table}
\begin{center}
 \begin{tabular}{||c c c||} 
 \hline
 Ionization Front & $\log \rm U_{\rm Max}$ & $\log \xi_{\rm Max}$ \\ [0.5ex] 
 \hline\hline
 H$^{0}$-H$^{+}$ & -0.98 & 0.62  \\ 
 \hline
 He$^{0}$-He$^{+}$ & -1.65 & -0.05  \\
 \hline
 He$^{+}$-He$^{++}$ & -0.48 & 1.12  \\
  [1ex] 
 \hline
\end{tabular}
\caption{Upper limits for the ionization parameter to have various ionization fronts.}
\label{T1}
\end{center}
\end{table}

properly describe the recombination process at
such densities.  {\tt cloudy} has such a model for one
and two electron ions \citep[section 3.2,
][]{Ferland17} but
not for many-electron systems.  CRM processes do
not affect the X-ray opacity since that is
produced by inner-shell electrons.  The models presented here
keep the total X-ray optical depth (at 1 keV) the same.  As
the density increases, H and He tend to become
more ionized due to the increasing contribution from collisional (CRM model), and the EUV and XUV
transmission increases. Because of the CRM,
changing the density is very similar to changing
the ionization parameter, due to due to the decreased efficiency in recombination. \citep{Ferland17}.

\subsection{Varying the Column Density}
\begin{figure*}
\centering
\includegraphics [width=\textwidth]{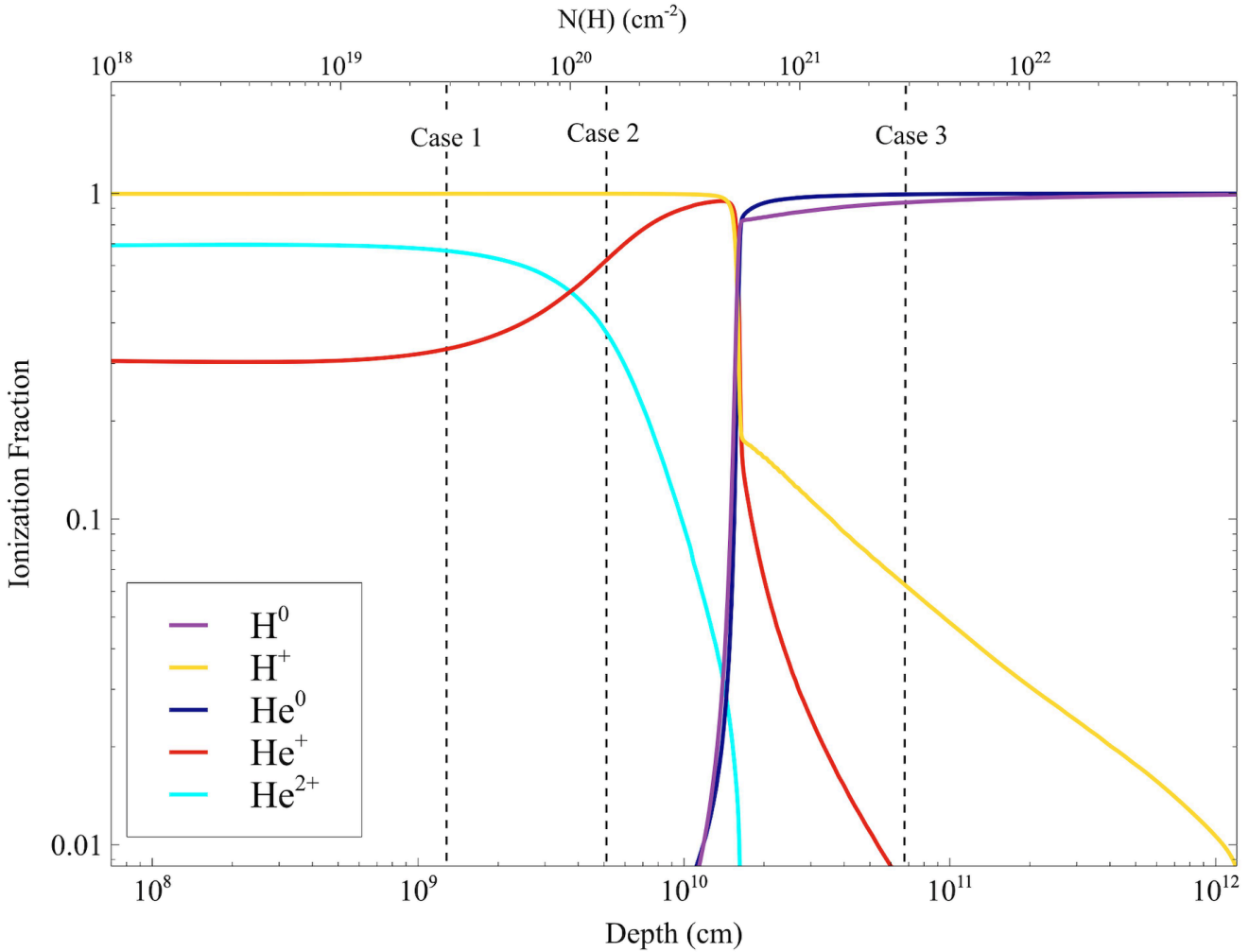}
 \caption{The ionization fraction of the noted ions as a function of the depth into the cloud (lower X-axis) in correspondence with the hydrogen column density (upper X-axis). The dashed lines show the depths corresponding to cases 1, 2, and 3, for assumptions of fixed gas density and incident ionizing photon flux. 
 Figure~\ref{fNHSED} shows how the transmitted and emitted SEDs would be for these three different depths.}\label{fIF}
\end{figure*}

The column density, N(H), is the number of atoms
along the line of sight per unit area. In the
simple case of a constant density n(H) cloud with
thickness \edit1{$\delta R$}, $\rm N(\rm H) = n(\rm H) \delta R$.  It is evident that
a thicker or denser cloud would have a larger column density. 
To see how different values of the obscurer’s
column density will affect the shape of the
transmitted/emitted SED, we first multiplied
optical depth of the LOS obscurer by four to make
the cloud thicker than what was observed. Figure~\ref{fIF} shows the ionization structure
versus depth for such a thick cloud, with an
ionization parameter of $\log\xi$=-1.2 erg cm
s$^{-1}$. In this Figure, the vertical dashed
lines show three different depths that we will
consider further. These are referred to as Case
1, Case 2, and Case 3 in the discussion below and
we will show that they produce transmitted SEDs
similar to those in Figure~\ref{fxi}.  These depths are
chosen because, as the Figure shows, they
correspond to different hydrogen and helium
ionization states.  The transmitted SED in each
case reproduces one of the three cases discussed
above and by D19b. 

As Figure~\ref{fIF} shows, the ionization
structure is very sensitive to the thickness of
the cloud. The absorbing power of a cloud is
proportional to the column density so larger
column densities produce lower ionization as averaged over the cloud.  

For the Case 1, both hydrogen and helium are
fully ionized. As the red line shows, in Case 1,
the amount of He$^{+}$ is much smaller than H$^{+}$ and He$^{++}$.  There is not enough
singly ionized helium to absorb the XUV. This means that the obscurer is transparent, and the
intrinsic SED is fully transmitted. We propose
that this is the case in most of AGNs: They DO
have disk winds, but if the winds are in a
transparent state no effects are observed. For
Case 2, He$^{++}$ recombines to form He$^{+}$
while H remains ionized. Both the transmitted SED
and emitted He II photons will ionize any
remaining H$^{0}$. This is why the He$^{+}$ zone
is also an H$^{+}$ zone. There is enough He$^{+}$ to block a portion of the XUV although a significant fraction is
transmitted.   The presence of a He$^{++}$-He$^{+}$ ionization front causes much
of the 54 eV$<h\nu <$ 200 eV transmitted SED to
be absorbed.  This reproduces Case 2 discussed by
D19b: the case in which the BLR holiday happens.
Finally, for Case 3, photons with
energies more than 13.6 eV are absorbed and
atomic H and He forms. This reproduces Case 3 of
D19b: a changing look quasar. In this case
H$^{+}$, He$^{+}$ and He$^{++}$ ionization fronts are present.

\begin{figure*}
\centering
\includegraphics [width=\textwidth]{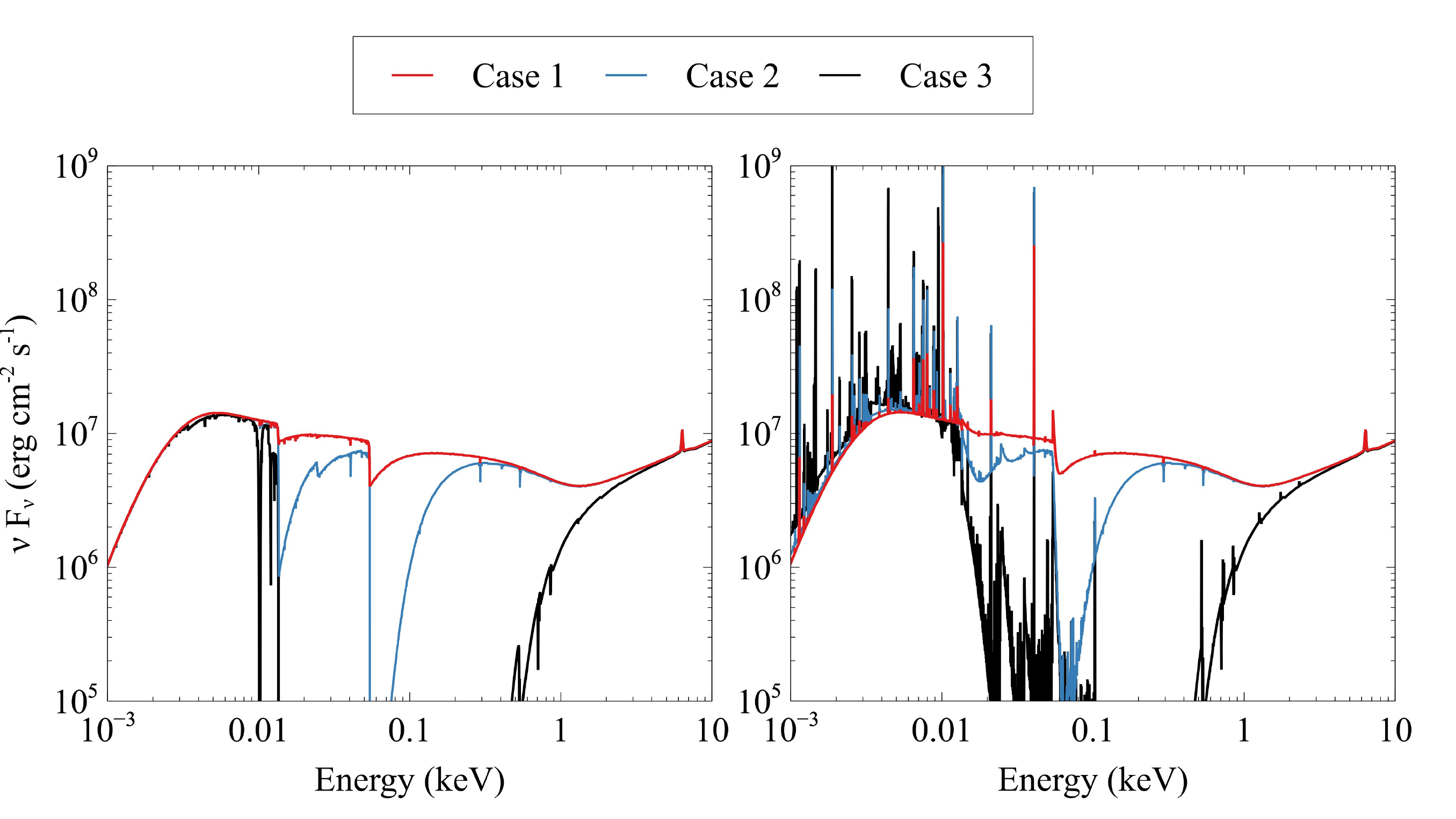}
 \caption{: Left panel: transmitted SED in
 different depths of a thick cloud. Right panel:
 total emission (sum of the transmitted and reflected continua plus attenuated incident continuum) from the obscurer for the same three
 cases.}\label{fNHSED}
\end{figure*}

Figure~\ref{fNHSED} shows the transmitted SED and
the emission from the cloud for the three cases.
The red line shows the SED for the Case 1 cloud.
This SED is not strongly extinguished since there
is no ionization front and hydrogen and helium
remain ionized. In Case 2 and Case 3 hydrogen and helium 
ionization fronts occur, the SED is strongly
absorbed and the diffuse continuum emission increases \citep{Korista01,Korista19, Law18}. 

\subsection{Varying the Hydrogen Density }

\begin{figure*}
\centering
\includegraphics [width=\textwidth]{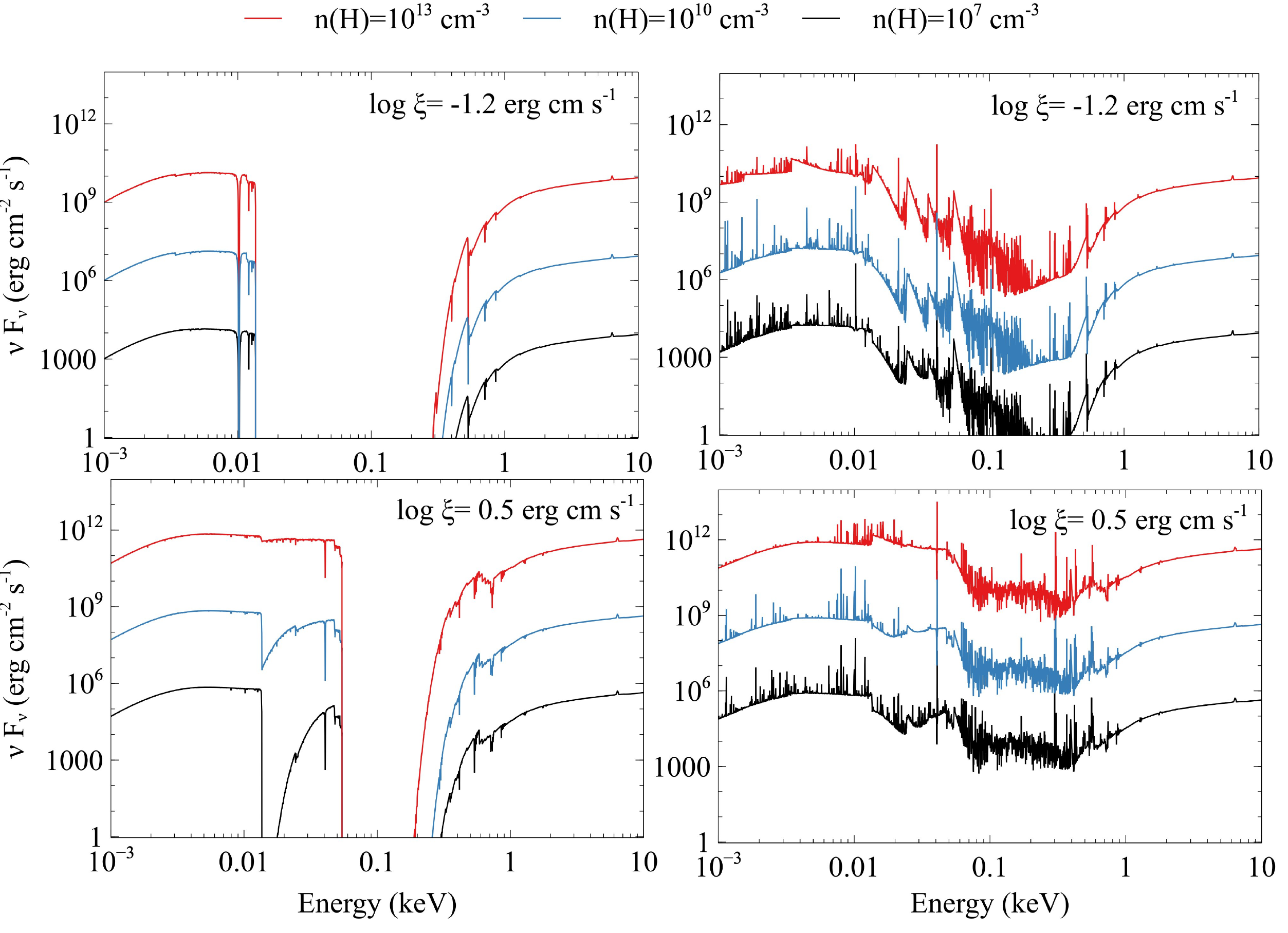}
 \caption{Changes in the SED for different values
 of the obscurer hydrogen density for two models
 of ionization parameters. Both upper panels have
 ionization parameter $\log\xi$=-1.2 erg cm s$^{-1}$. Upper left
 panel shows the results for the transmitted SED
 and the upper right shows the results for the
 total emission from the obscurer (sum of the transmitted and reflected continua plus attenuated incident continuum). Lower panels show the
 same concept for and ionization parameter
 $\log\xi$=0.5 erg cm s$^{-1}$.}\label{fnHSED}
\end{figure*}

The hydrogen density of the LOS obscurer is poorly
constrained. Here, we predict SEDs for different
hydrogen densities while keeping the soft X-ray
optical depth constant. We investigate the effects
of changing density for two different values of
the ionization parameter ($\log\xi$=0.5 erg cm
s$^{-1}$ corresponding to Case 2, and
$\log\xi$=-1.2 erg cm s $^{-1}$ , corresponding to
Case 3). Varying $\xi$ is equivalent to changing
the distance between the obscurer and the
accretion disk.  Figure~\ref{fnHSED} shows the results. 

In each of the four panels of the Figure~\ref{fnHSED}, the ionization parameter is
kept constant while the density varies. As noted
earlier, simple homology relations suggest that
clouds with similar ionization parameters, but
different densities and flux of ionizing photons,
should have the same ionization \citep{Ferland03}.
The results for the lower ionization clouds
are fairly similar.  The results for the higher
ionization parameter shown in the lower-left panel
are surprising because the higher density clouds
are more highly ionized and transparent. This is
caused by the increasingly important role of
collisional ionization from highly excited states in the
high-radiation environment, as discussed earlier.

As the right panels of Figure~\ref{fnHSED} show,
for energies in UV/optical regions, the obscurer
also emits. This extra emission is stronger when
the obscurer is denser. This excess of emission is
mainly due to hydrogen radiative recombination in
the optical and NIR and bremsstrahlung  in the IR.  

\begin{figure*}
\centering
\includegraphics [width=\textwidth]{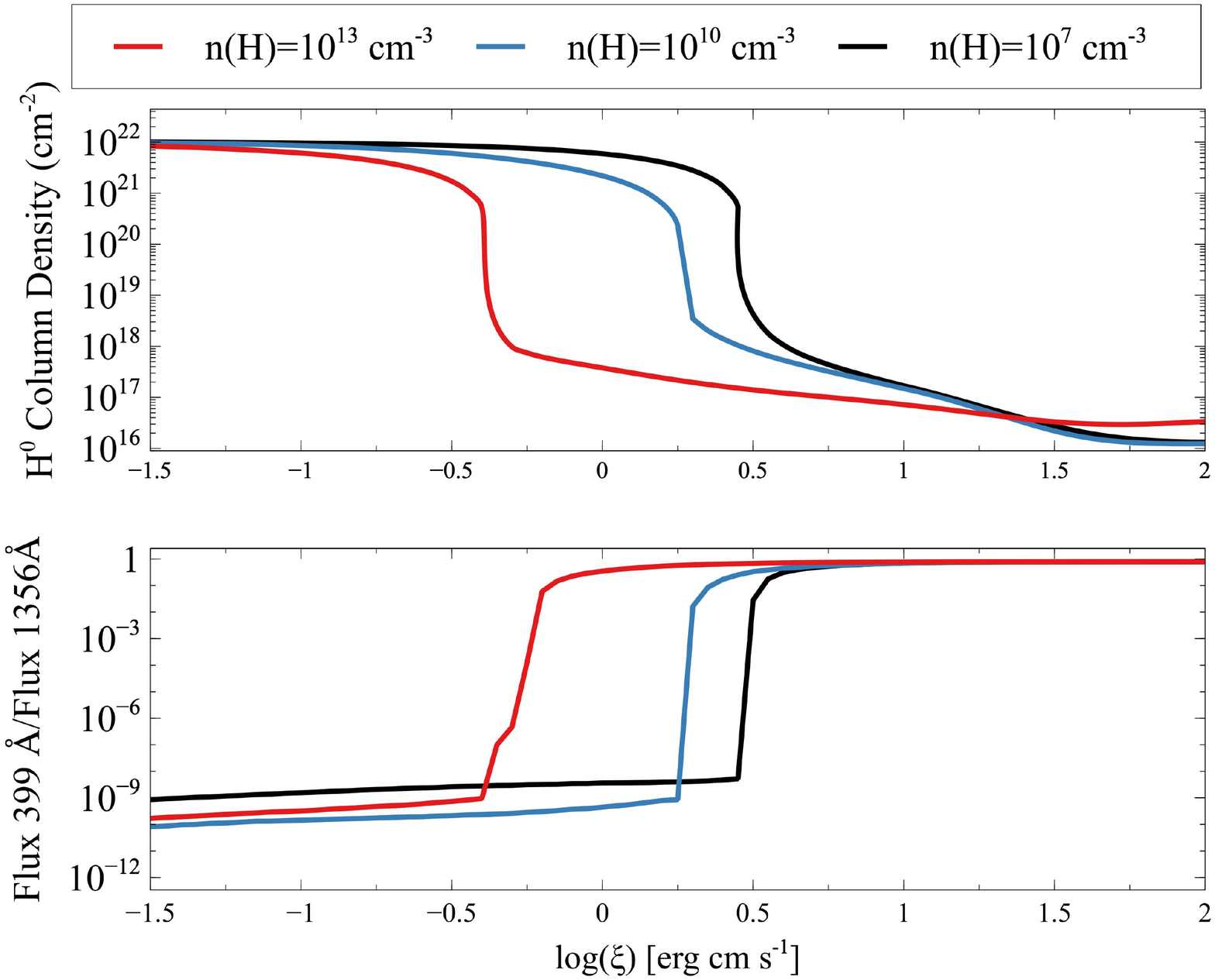}
 \caption{Top panel shows the variation of atomic hydrogen column density as the ionization parameter changes, for three different hydrogen densities. The bottom panel shows the ratio of 399\AA/1365\AA \ as a function of the ionization parameter for the same three hydrogen densities.}\label{fnHratio}
\end{figure*}

Figure~\ref{fnHratio} is similar to Figure~\ref{fxiRatio}. In this Figure, the upper panel shows the
variations of the H$^{0}$ column density as a function
of the ionization parameter for three different
hydrogen densities. As in Figure~\ref{fxiRatio}, the lower
panel shows the ratio of the intensity of the
transmitted continuum at 399 \AA \ (EUV) relative to
that at 1356 \AA \ (FUV), this time for three
different values of hydrogen density.

Please note that for the low densities, the CRM
effects discussed earlier, are not important and
simple photon conserving arguments hold. However,
the CRM effects are considerable for higher
densities and would affect the ionization front
algorithm. This is the reason that we see
different hydrogen ionization fronts for different
values of the hydrogen density when the ionization
parameter is $\log \xi$=0.5 erg cm s$^{-1}$.

\subsection{Varying the Metallicity}
We have assumed solar metallicity so far, and we
next vary this. Since most soft X-ray absorption
is produced by K-shell electrons of the heavy
elements, one can expect that if we raise the
metallicity by some factor, the hydrogen column
density will fall by the same factor to keep the
heavy element column density, and X-ray optical
depth, the same. We checked this by modeling the
transmitted and emitted SEDs for two very
different values of the metallicity in two
different models of ionization parameters.

\begin{figure*}
\centering
\includegraphics [width=\textwidth]{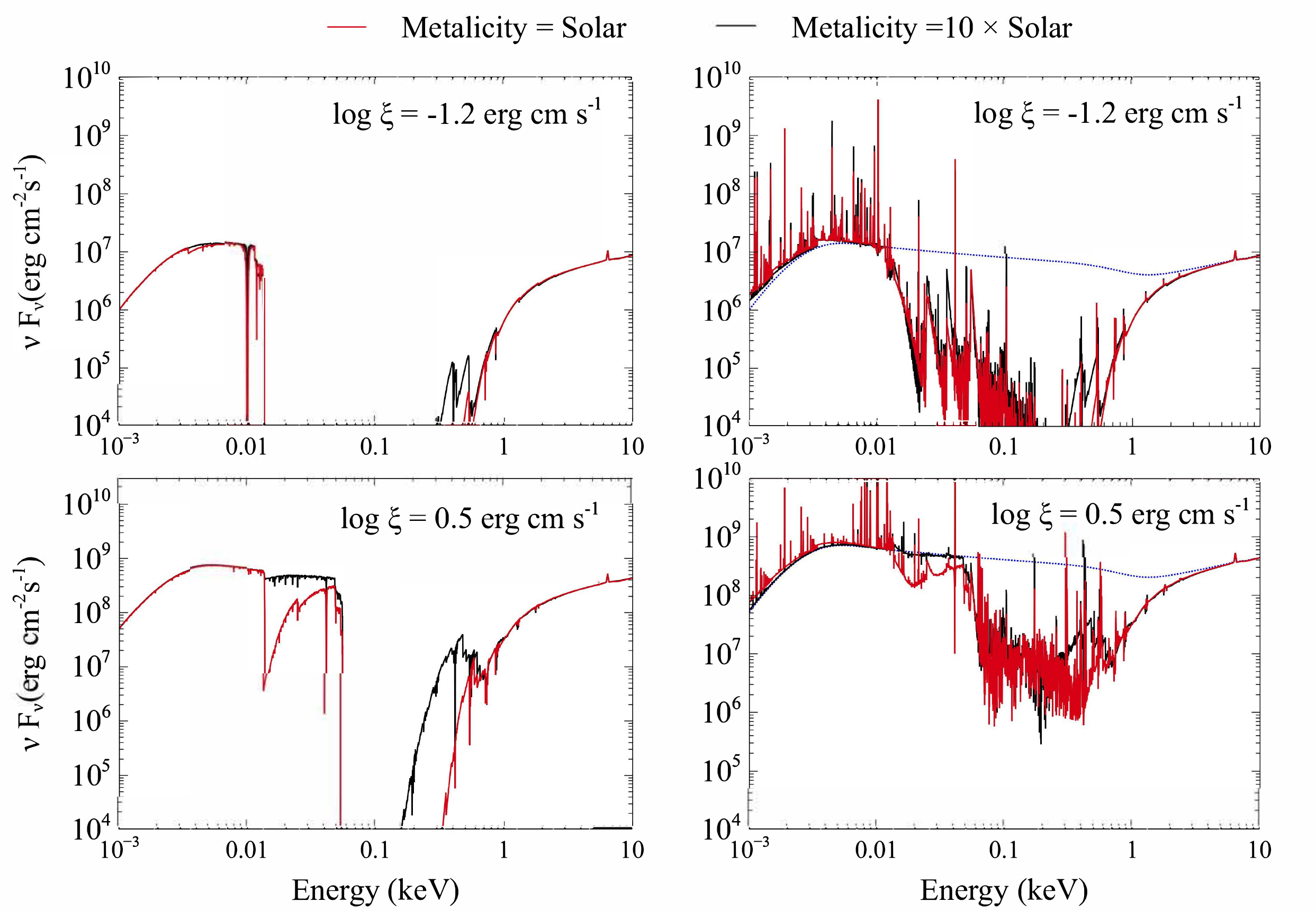}
 \caption{Changes in the SED for different values
 of the metallicity for two models
 of ionization parameters. Both upper panels have
 ionization parameter $\log\xi$=-1.2 erg cm s$^{-1}$. Upper left
 panel shows the results for the transmitted SED
 and the upper right shows the results for the
 total emission from the obscurer (sum of the transmitted and reflected continua plus attenuated incident continuum). Lower panels show the
 same concept for an ionization parameter
 $\log\xi$=0.5 erg cm s$^{-1}$.}\label{fmetSED}
\end{figure*}

The upper panels of Figure~\ref{fmetSED}  show the
transmitted SEDs (upper left) and the emission
from the obscurer (upper right) for solar metallicity ($\rm Z_{\odot}$) and
10$\times$ $\rm Z_{\odot}$ for the ionization
parameter $\log\xi$=-1.2 erg cm s$^{-1}$. 
Clearly, the transmitted SEDs of the upper left
panel fall in Case 3 category. As
both upper panels show, changing the metallicity
does not profoundly affect the SED when we are
in this case.  

The lower panels have an intermediate ionization
parameter, $\log \xi$=0.5.  Much of the soft X-ray
extinction in the lower-left panel is produced by
inner shell photoabsorption of the heavy elements.
Although the 1 keV extinction, mainly produced by
inner shells of O, C, is the same, the extinction
around 200 – 400 eV changes significantly.  The
opacity in this range is mainly due to He and H
(figure 10 of D19a) and little H$^{0}$ or He$^{0}$ is
present.  There is almost no hydrogen ionization
front in the Z=10$\times$Z$_{\rm \odot}$ case because the
hydrogen column density is ten times smaller, as
shown next.  So, in this respect, the model
behaves like Case 2 with no hydrogen ionization
front.

\begin{figure*}
\centering
\includegraphics [width=\textwidth]{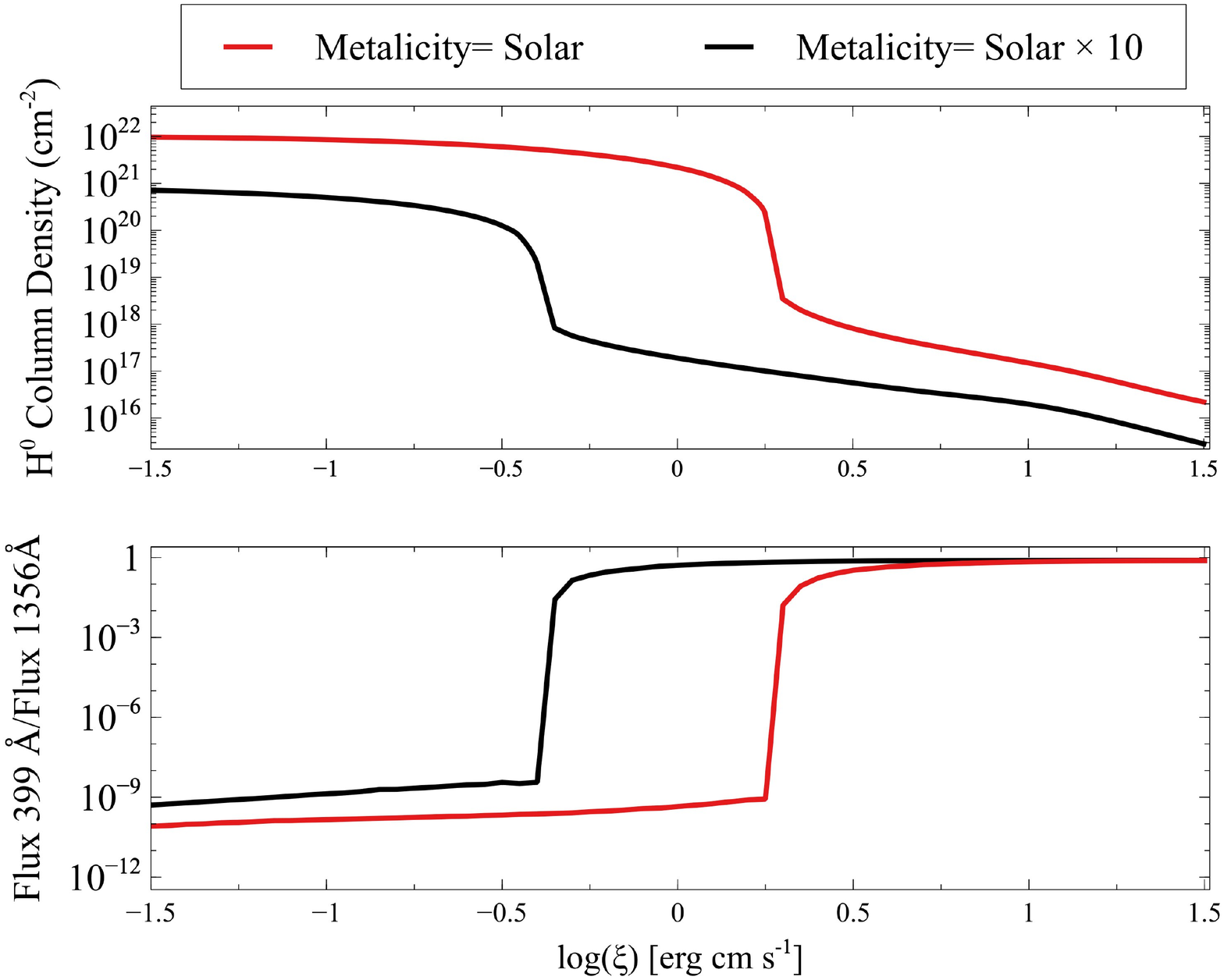}
 \caption{Top panel shows the variation of the
 obscurer atomic hydrogen column density as the
 ionization parameter changes for two different
 values of the metallicity. The bottom panel shows
 the ratio of the 399\AA/1365\AA\ transmitted continuum as a function of the ionization parameter. }\label{fmetRatio}
\end{figure*}

Figure~\ref{fmetRatio} is the equivalent to Figure~\ref{fnHratio}, but for different
metallicities.  As expected, for the case with higher metallicity, the hydrogen column
density is about 10 times smaller at low and high
ionization parameters.  However, there are much
greater differences at intermediate $\xi$,
around $\log\xi=\pm $0 erg cm s$^{-1}$.  For these
parameters, the locations of the hydrogen and
helium ionization fronts, in the high-Z case,
straddle the outer edge of the cloud and large
changes in opacity occur.

\subsection{Summary }
All of the above discussions show that the SED
filtered through an obscurer or reflected by it
might change slightly or dramatically, depending
on the obscurer’s properties. When the filtering
happens, as in the case of NGC 5548, we could
observe the original SED, the absorbed SED, and
the reflected SED. Usually, it is not possible to
directly measure the obscurer’s properties,
however, by having the SEDs transmitted/reflected
through the obscurer, one can find the
characteristics of the obscurer itself, as demonstrated in D19b \& D20a. 

\section{LIMITS ON THE OBSCURER:  THE GLOBAL COVERING FACTOR}
While the obscurer’s properties such as its hydrogen
density and ionization parameter play a pivotal
role in determination of the shape of the
transmitted/reflected SED, we still need to know
another characteristic of the obscurer, its global
covering factor, to determine the geometry of the
obscurer. 

As seen from Earth, the obscurer in
NGC 5548 absorbed a great deal of the energy
emitted by the AGN since the ionizing photon
luminosity fell from 2$\times$ 10$^{44}$ erg s$^{-1}$ to roughly
7.7$\times$10$^{43}$ erg s$^{-1}$ after
obscuration (based on the SED modeling of
\cite{Mehd16}). The obscurer removed roughly
11.2$\times$ 10$^{43}$ erg s$^{-1}$, as seen by us if it fully
covers the continuum source. Energy is conserved,
so this light must be reradiated. 

The total energy absorbed and emitted by the
obscurer is determined by its global covering
factor (explained in section 2.2), which is
unknown.  However, we do know that it persists for
several orbital timescales \citep{Kaastra14}
so the geometry may be something like a symmetric
cylinder covering 2$\pi$ around the equator (Figure1 of D19b).  This suggests that the GCF may be
significant. To check this, we derive the
emission-line luminosities predicted by the LOS
obscurer model to examine its spectrum and
establish an upper limit on the LOS obscurer
global covering factor. 

\subsection{The Luminosity of the Obscurer}
Here we vary the distance between
the obscurer and the central black hole (R$_{\rm obs}$) to
judge the magnitude of the effect on the obscurer’s emission lines.
The most significant impact of doing this is to
change the ionization parameter since it is
proportional to R$_{\rm obs}^{-2}$.
The location of the wind in NGC 5548 is fairly well known.
There is spectroscopic evidence showing the wind is located between the BLR and the source (i.e. R$_{\rm obs}<10^{16}$cm). 
However, this is not the case for all AGNs, so it is informative to show how 
variations of the distance  affect some of the wind's properties to establish
a diagnostic for other studies. 
Here we illustrate a novel method that uses limits to line intensities 
to establish bounds on the global covering factor and the location of the wind. 

\begin{figure*}
\centering
\includegraphics [width=\textwidth]{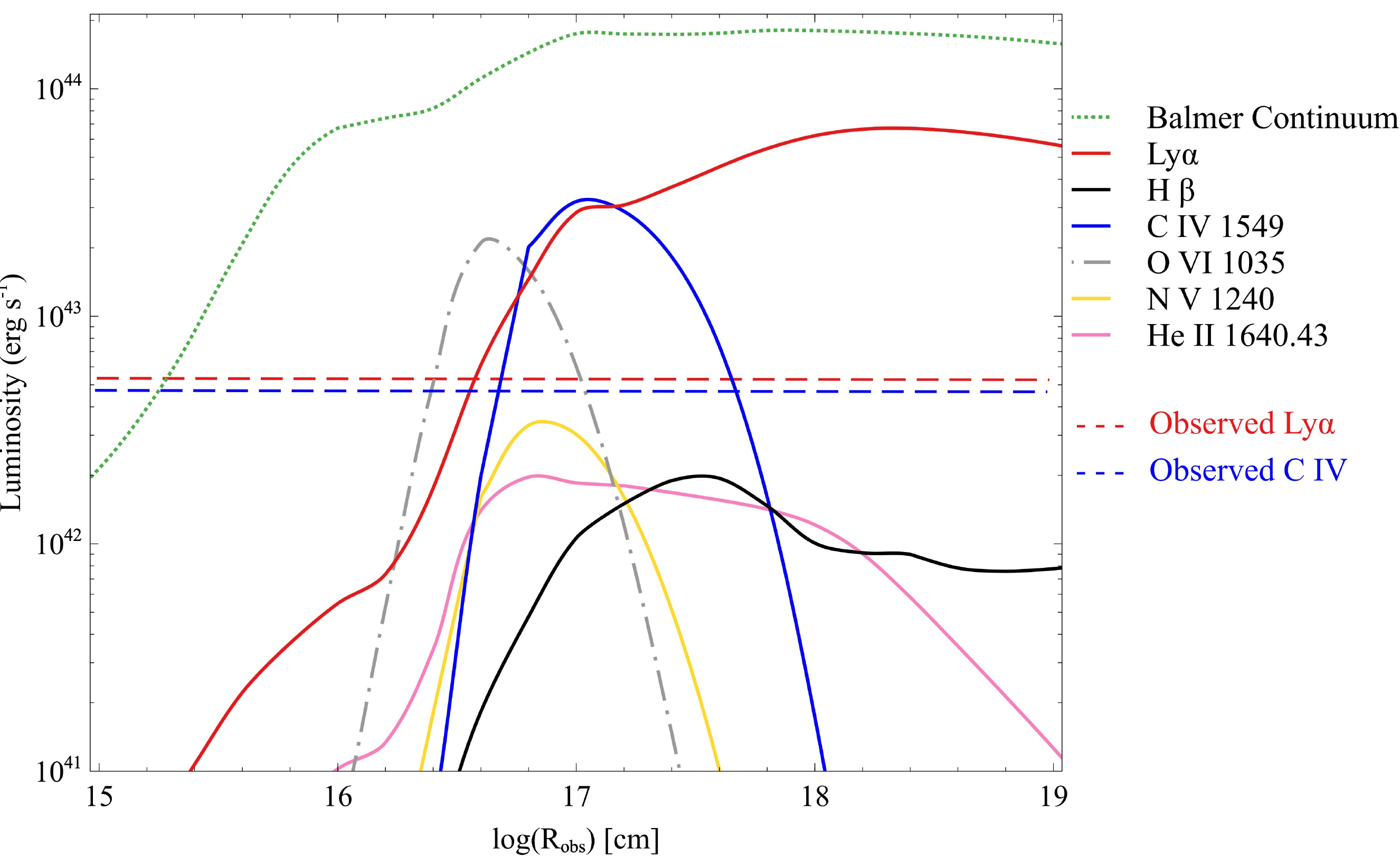}
 \caption{Changes in the luminosity of the
 emission features of the obscurer as the
 separation between the obscurer and the black hole varies
 between 10$^{15}$ and 10$^{19}$cm.  The 
 luminosities are predicted for full source coverage. The observed values refer to a combination of broad, medium broad, and very broad emission components observed by HST.} \label{fobslum}
\end{figure*}

Figure~\ref{fobslum} shows 
predicted emission-line luminosities for an
obscurer that fully covers the continuum source, a
GCF of 100$\%$ and has a hydrogen density of 10$^{10}$
cm$^{-3}$. To create the model we considered a constant optical depth at 1 keV.  

The predicted luminosities in Figure~\ref{fobslum}
can be compared with observations to obtain an upper
limit to the obscurer’s GCF. The range of radii corresponds to changing the
ionization parameter by a factor of 10$^8$. The strongest
spectral features are the Balmer continuum, Ly$\alpha$, \civ, and O VI.  Of these, the Balmer continuum and Ly$\alpha$\  have the least dependence on the unknown radius.   We can compare the observed luminosity of Ly$\alpha$\  with these predictions to set an upper limit on the obscurer’s GCF.

Table 1 of \cite{Kriss19} lists the UV emission lines in NGC 5548. Based on this information, the observed flux (broad+medium broad+very broad)
for Ly$\alpha$\  is 8.14$\times$10$^{-12}$ erg
cm$^{-2}$ s$^{-1}$ (after correction for foreground Milky Way Galaxy extinction), and for the luminosity distance given
by \cite{Mehd15}, it has a luminosity of
5.36$\times$10$^{42}$ erg s$^{-1}$. This value is
shown with a red dashed line in
Figure~\ref{fobslum}.  The predicted Ly$\alpha$\ luminosity can be
either smaller or larger than this value.  The Figure shows that we predict a Ly$\alpha$\  luminosity of $\sim$ 5.46$\times$10$^{41}$ erg
s$^{-1}$ for R$_{\rm obs}$= 10$^{16}$ cm and full
coverage.  This value is almost 10 times smaller
than that observed. It means that, based on
the Ly$\alpha$\  luminosity, if the obscurer is located
at 10$^{16}$ cm from the source, there is no need to constrain its
GCF. If the same obscurer is located farther away, at
R$_{\rm obs}$= 10$^{17}$ cm for instance, the
predicted value is $\sim$ 2.85$\times$10$^{43}$ erg
s$^{-1}$. This limits the obscurer global covering
factor to be less than $\sim$ 19$\%$ . This shows
that there is an interplay between the location of
the obscurer, the emission line luminosities, and its
global covering factor.

The \civ\  line is much more model sensitive due to
its dependence on the ionization parameter. Based
on table 1 of \cite{Kriss19}, the \civ\  flux (broad+medium broad+very broad) is 7.17$\times$10$^{-12}$ erg
cm$^{-2}$ s$^{-1}$ (after correction for foreground Milky Way Galaxy extinction), and this leads to a luminosity of 4.72$\times$10$^{42}$ erg
s$^{-1}$.  Figure~\ref{fobslum} shows that the
predicted emission line luminosity varies by many orders of
magnitude.  For R$_{\rm obs}$= 10$^{16}$ cm, the
predicted \civ\  line luminosity, is much smaller
than the observed value, as indicated by the blue dashed
line. This means that GCF can be as large as 100$\%$.  For larger
radii, corresponding to smaller ionization
parameters, the luminosity increases, reaching a maximum \civ\  luminosity of 3.19$\times$10$^{43}$ erg
s$^{-1}$, almost 7 times brighter than the observed
value, requiring a covering factor less than one.
We come away with the picture that the emission
from the obscurer can be a contributor to the
observed broad emission, and could, in fact,
account for all of it, depending on the location
of the obscurer. Next we investigate how the emission from the obscurer varies as a function of both its location and hydrogen density, for the three different cases discussed earlier.

\section{EMISSION FROM THE WIND}

D19b proposed that changes occuring in the base of the disk
wind, the equatorial obscurer, explains the BLR
holiday. This obscurer is located close to the
central source and is assumed to absorb a
significant amount of the SED striking the BLR. To
conserve the energy, the obscurer must re-emit
this energy. D20 showed that the equatorial
obscurer produces its own emission lines and came
up with a model in which the obscurer is
not a dominant contributor to most of the
strong emission lines, while it can be considered
as the main He II and Fe K$\alpha$ emission source. In their model, the emission lines observed are indeed a combination of a broad core (produced in the BLR) and a very broad base (produced by the equatorial obscurer). 
Below we investigate various emission lines
produced by the equatorial obscurer in each of the three cases
that were discussed earlier. 

\subsection{Very broad emission lines}
\begin{figure*}
\centering
\includegraphics [width=\textwidth]{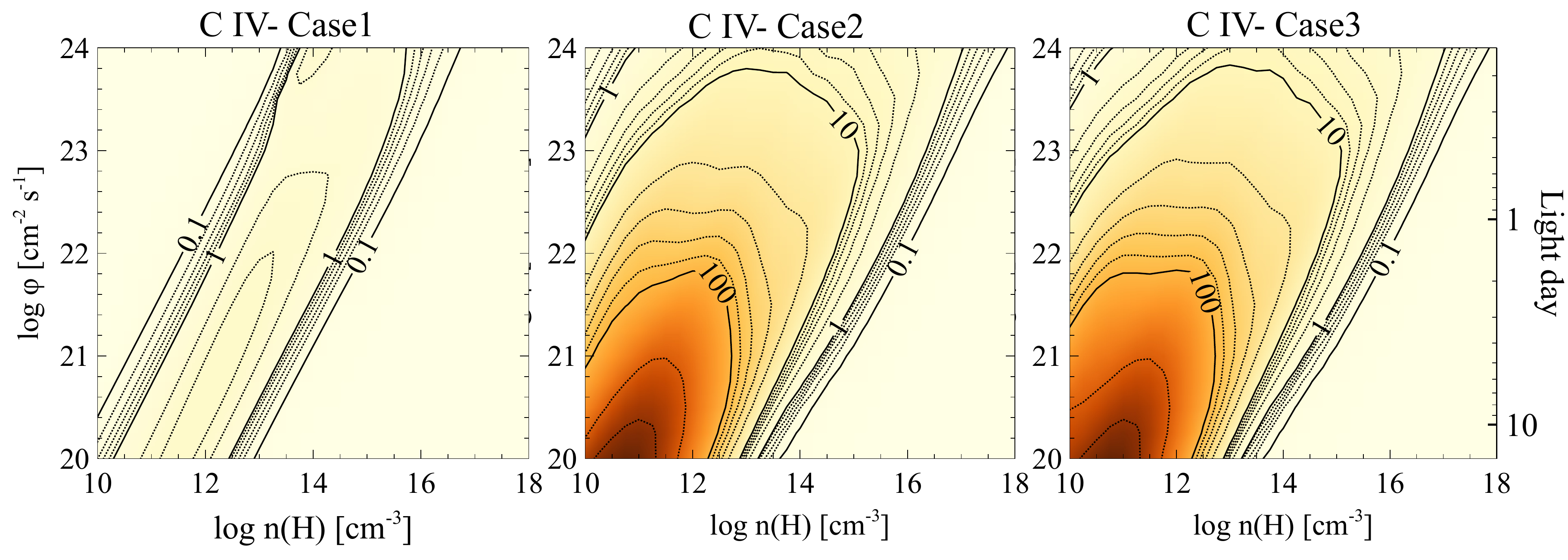}
 \caption{The EW of \civ\ emission line as a function of both the flux of hydrogen ionizing photons and the hydrogen density. Different panels show the variation of the EW for each of the discussed cases. The intervals between decades are logarithmic in 0.2 dex steps.}\label{fciv}
\end{figure*}

Figures~\ref{fciv}\ to \ref{fhb} illustrate emission line
equivalent widths for a variety of lines.  Each of these Figures belongs to a single emission line and shows its behavior for the
obscurer in Case 1, 2, or 3. To produce
three different Cases, the optical depths are
chosen such that each model of the obscurer
falls in the middle of each region of figure 4 in
D19b. This leads to a typical example of an
obscurer for each case. 

In each panel, the EW of the emission line is
modeled as a function of both the flux of photons
produced by the source ($\phi(\rm H)$) and the hydrogen
density. To create these models, we used the SED
of \cite{Mehd15} in {\tt cloudy} (developer
version), while we assumed photospheric solar
abundances \citep{Ferland17}. We produced
two-dimensional grids of photoionization models as
we have done in D20. Each grid includes a range of
total hydrogen density, 10$^{10} \rm cm^{-3} < n(\rm H) <$ 10$^{18}\rm cm^{-3}$, and a range of
incident ionizing photon flux, 10$^{20}$ s$^{-1}$ cm$^{-2} <
\phi(\rm H)<$ 10$^{24} \rm s^{-1} \rm cm^{-2}$. The flux of ionizing
photons, the total ionizing photon luminosity
Q(H), and the distance in light days are related
by:

\begin{equation}
    \phi(\rm H)=\frac{\rm Q (\rm H)}{4\pi r^2},
\end{equation}
which indicates when the obscurer is closer to the
source (smaller r), it will receive a larger
amount of ionizing photon flux (larger $\phi(\rm H)$).

Considering the \civ\  lag and based on Figure 4 of D20a, the \civ-forming region of the BLR has an incident ionizing
photon flux of approximately  10$^{21}\rm cm^{-2}\rm s^{-1}$, while
the equatorial obscurer has an incident ionizing flux of almost
10$^{22.5}\rm cm^{-2}\rm s^{-1}$, which means the
equatorial obscurer is $\sim$6 times closer to the
central source than a typical point in the BLR.
Such an obscurer emits lines with a FWHM $\sim$4 time
broader than the BLR if motions are virialized. As
proposed by D20, the line EWs observed by \hst\  and
other space telescopes are a combination of the
broad emission from the BLR and the very broad
emission from the equatorial obscurer. 

\begin{figure*}
\centering
\includegraphics [width=\textwidth]{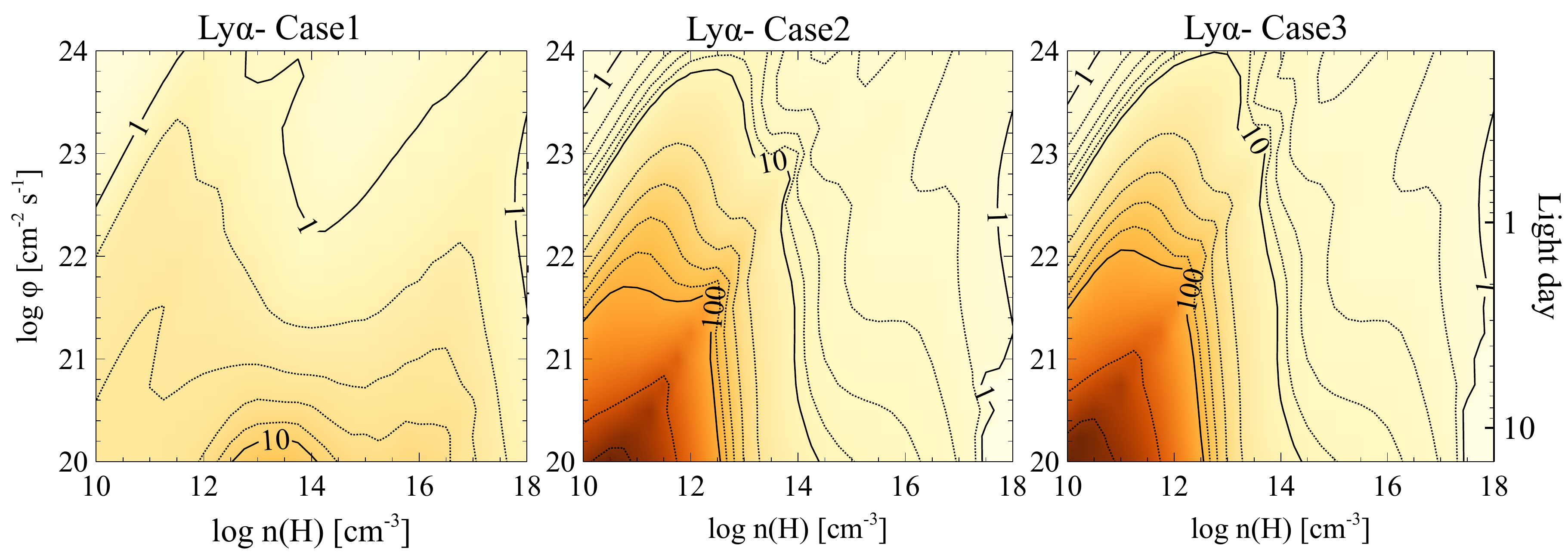}
 \caption{The EW of Ly$\alpha$\  emission line as a function of both the flux of hydrogen ionizing photons and the hydrogen density. Different panels show the variation of the EW for each of the discussed cases.}\label{flya}
\end{figure*}

\begin{figure*}
\centering
\includegraphics [width=\textwidth]{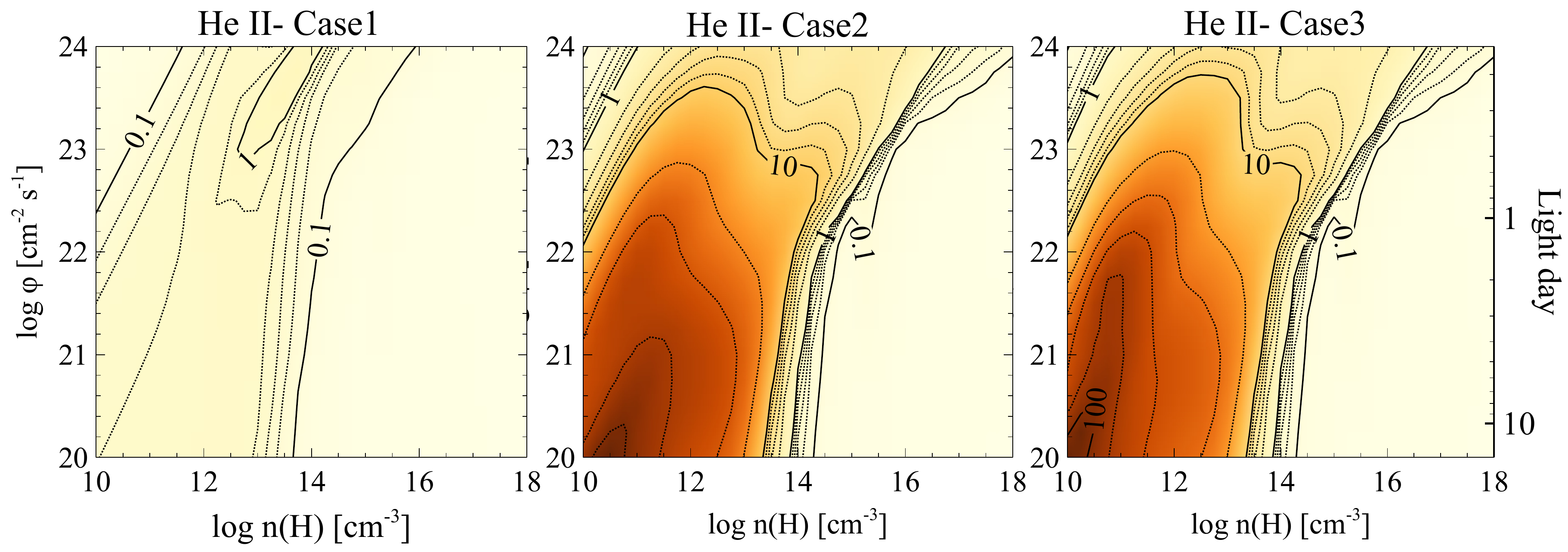}
 \caption{The EW of \heii\  emission line as a function of both the flux of hydrogen ionizing photons and the hydrogen density. Different panels show the variation of the EW for each of the discussed cases.}\label{fheii}
\end{figure*}

\begin{figure*}
\centering
\includegraphics [width=\textwidth]{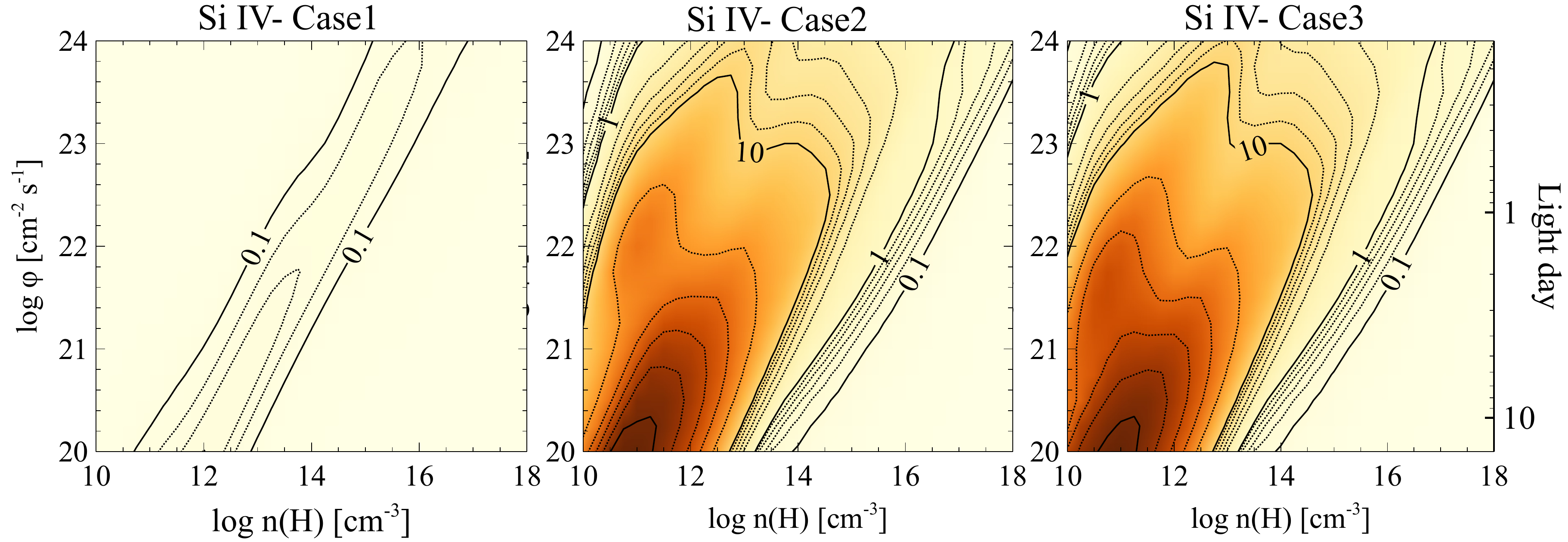}
 \caption{The EW of \siIV\ emission line as a function of both the flux of hydrogen ionizing photons and the hydrogen density. Different panels show the variation of the EW for each of the discussed cases.}\label{fsiiv}
\end{figure*}

\begin{figure*}
\centering
\includegraphics [width=\textwidth]{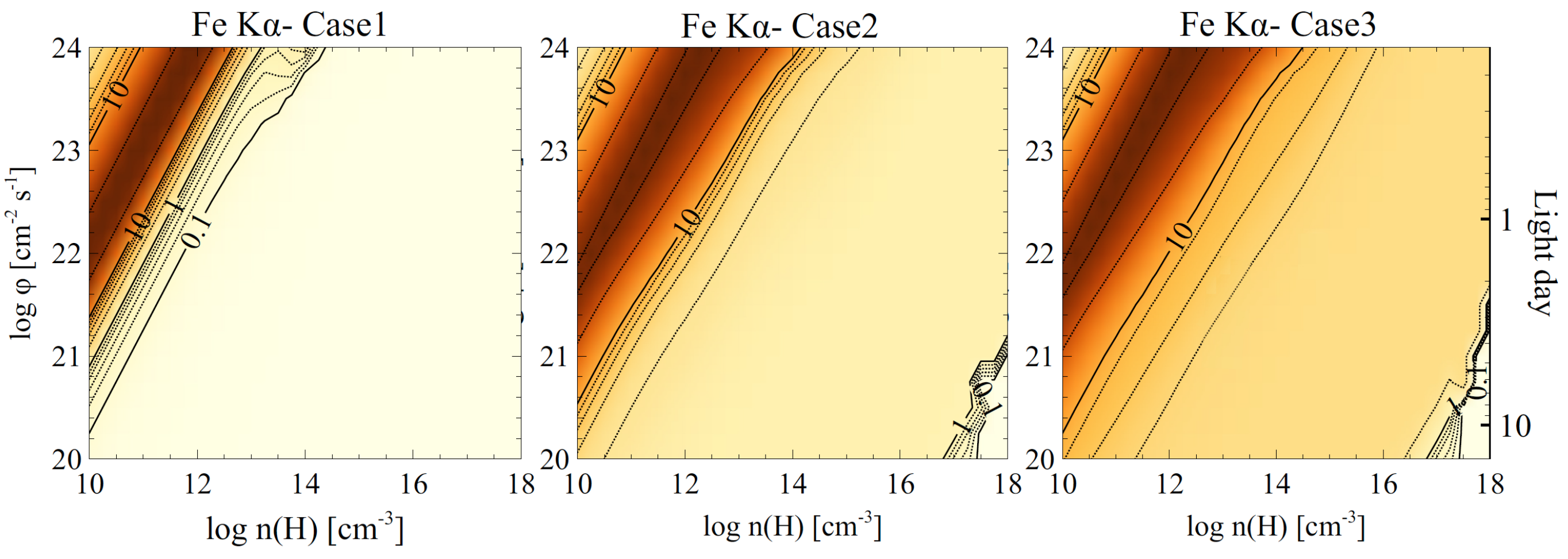}
 \caption{The EW of Fe K$\alpha$\ emision line as a function of both the flux of hydrogen ionizing photons and the hydrogen density. Different panels show the variation of the EW for each of the discussed cases.}\label{feka}
\end{figure*}

\begin{figure*}
\centering
\includegraphics [width=\textwidth]{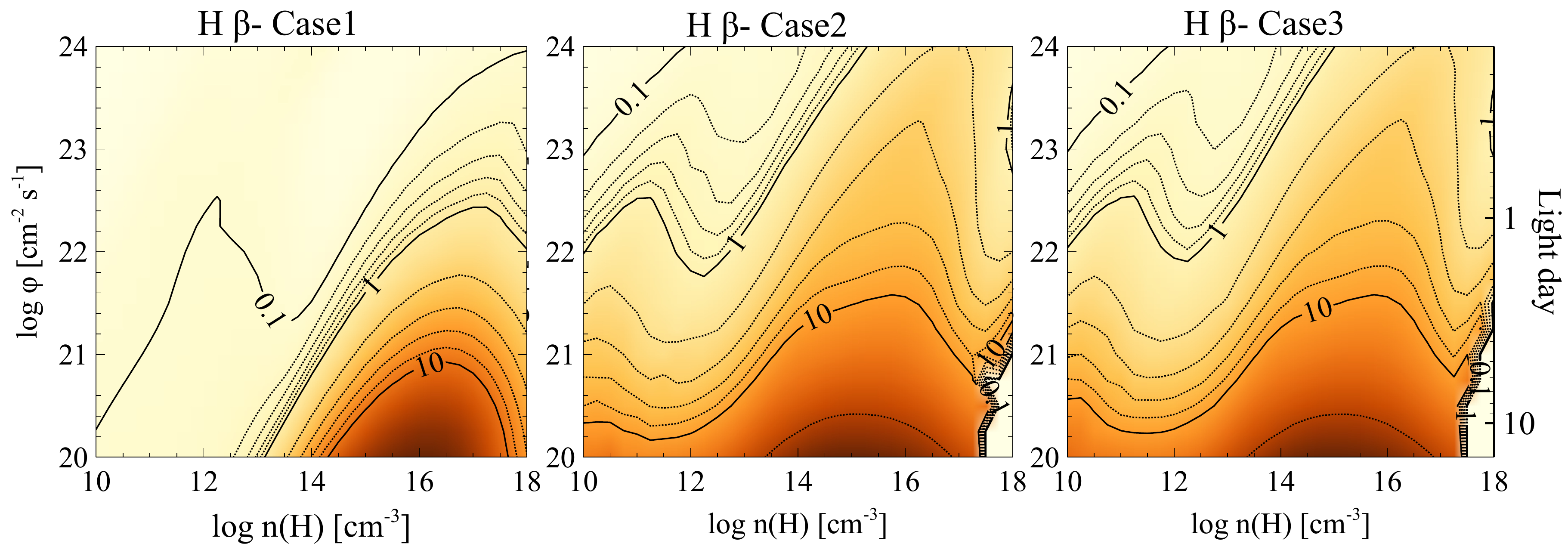}
 \caption{The EW of \hb\ emision line as a function of both the flux of hydrogen ionizing photons and the hydrogen density. Different panels show the variation of the EW for each of the discussed cases.}\label{fhb}
\end{figure*}

As Figures~\ref{fciv} to \ref{fsiiv} show, when the wind is in a transparent state (Case 1), it emits
very small amounts of \civ, \heii, Ly$\alpha$
and \siIV.  In such a situation, almost all of the
observed (broad+very broad) emission lines are
mainly broad emission produced by the BLR and the
equatorial obscurer has almost no contribution,
regardless of where it is located or what its
density is. This may be the case in most AGNs. 
However, when in Case 2 state, the
contribution of the obscurer becomes significant. This
means that by transformation from Case 1 to Case 2,
there will be a change in the observed EW of the
mentioned lines, which may lead to a holiday.

The predictions are different for Fe~K$\alpha$ and \hb\  (Figures~\ref{feka} and \ref{fhb})
emission lines. A transparent disk wind which is
located close enough to the central source
($\phi(\rm H)>10^{21}\rm s^{-1}\rm cm^{-2}$) with
a relatively low density ($\rm n(\rm H)~<10^{12} \rm
cm^{-3}$)  (Figure~\ref{feka}, Case 1, upper left
corner) will produce Fe~K$\alpha$ as much as a translucent
disk wind. In this regime, the ions are too highly
ionized to permit the Auger effect. Meanwhile, line
photons are still affected by resonant scattering.
There is no destruction mechanism so they can
leave the disk. As a result, Fe K$\alpha$ XXV and  XXVI will be emitted at 6.67 keV and 6.97 keV \citep{Reynolds03}. In this
case, and based on the discussion of D20, both
transparent and translucent disk winds can be a
major contributor to the observed Fe~K$\alpha$ emission.
For NGC 5548, this confirms that the total amount
of observed Fe~K$\alpha$ will be almost constant
before, during or after the holiday. Indeed, it does not matter which Case
the wind state is in, it will always produce roughly
the same amount of Fe~K$\alpha$. 

A transparent wind which is located far enough from the source ($\phi(\rm H) < 10^{22}\rm s^{-1}\rm cm^{-2}$) with a large enough density ($n(\rm H)>10^{14} \rm
cm^{-3}$) would produce almost as much \hb\ as
produced by a translucent wind (Figure~\ref{fhb},
Case 1, lower right corner). Table 6 of
\cite{Pei17} reports that the flux of \hb\  had the
smallest change (6$\%$) among the other strong
emission lines (\civ, \heii, Ly$\alpha$\  and
\siIV). This is consistent with the behavior of \hb\  in our model, in which by transforming the
wind from Case 1 (non-holiday) to Case 2 (holiday)
the amount of \hb\  emitted by the equatorial
obscurer is not affected dramatically, however, as we mentioned earlier, it is still affected enough to be consistent with the observed emission-line holiday.
Depending on the optical depth, a translucent wind
can be considered as Case 2 or Case 3. As Figures~\ref{fciv}\ to \ref{fhb}\ show, for all of
the emission lines, the behavior of a Case 2
obscurer is very similar to that of Case 3. D20
explained that a Case 3 wind could be a contributor
to the changing-look phenomenon.

\subsection{Summary}
The equatorial obscurer is located close to the
central source and, depending on whether it is
transparent or translucent, it absorbs a small or
large portion of the SED striking the BLR. The
obscurer conserves energy by emitting very broad
emission lines \edit1{which are explained and investigated by D20.}. For several of
the strong UV emission lines, when the obscurer
transforms from normal to a holiday phase, there is
an increase in the very broad component, since the
obscurer absorbs more energy than before. At the
same time, the EWs of BLR broad emission components
decrease due to receiving less energy from the
source. The combination of these two phenomena
leads to a decrease in the total (broad + very
broad) observed emission line, i.e. a holiday. For some other lines, such as Fe K$\alpha$ and \hb, the
obscurer always emits almost the same amount of
very broad emission line. This means its phase will
not have any effects on absorbing that specific energy. As a result, it will not have a considerable
effect on the BLR energy, and the total observed EWs of these lines seem almost constant during the
transformation from normal to the holiday, and vice
versa. 

\section{DISCUSSION: THE ROLE OF DISK WINDS IN THE AGN PHENOMENON }

The STORM campaign showed that obscuration due to
disk winds plays a major role in AGN variability. 
Obscurers are part of a disk wind so they can be
common. These obscurers could cover a significant
part of the continuum source and thus alter the
SED striking the BLR or absorption clouds. As we show
in Figures~\ref{fxi} and \ref{fnHSED}, there are
conditions for which the obscurer is almost transparent, and
so has no effect on the transmitted SED. We propose
that this is the case in most AGNs in which
no holiday is observed. If their effect is not
observed, then it could be that they are in
the transparent state (Case 1) which could be a
result of low densities (D19b) or a high ionization
parameter. 

There are fewer UV reverberation campaigns than optical studies
and holidays similar to NGC 5548 would be difficult to detect with optical data alone. 
In NGC 5548 the H$\beta$ EW did not change greatly during the holiday,
(the flux deficit was only 6$\%$ \citep{Pei17})
and our calculations predict that H$\beta$  does not change dramatically 
when the obscurer varies between Case 1 and Case 2.
This makes it difficult to detect a holiday with optical data alone.
Holidays may be more common than we now suspect.

When the obscurers are dense enough or highly
ionized, they will emit. This emission could be a
source of non-disk emission which will contribute
to the observed broad emission.  It is also
possible that what is absorbed is radiated in FUV
ranges, so is not detectable. 

In such dense cases, the obscurer is removing a
great deal of energy, and this can lead to the
absence of BLR emission lines (D19b). This
situation may provide an alternative cause
of the changing-look phenomenon in AGNs. Right now,
the only explanation for this phenomenon is that
the source gets faint, so the BLR lines disappear.
The obscurer can cause the changing look phenomenon
in cases when the source is still bright, but the BLR
lines are gone.

The \storm\  campaign discovered an unexpected relationship
between the ionizing SED and the response of the spectral lines.
The extensive physical simulations carried out in this and previous papers 
quantifies how an intervening translucent screen
can modify the SED and produce what \hst\  and \xmm\  observed for NGC 5548. 
This screen is most likely the inner portions of a
disk wind so the \hst spectral observations
provide an indirect probe of a phenomenon that
cannot be directly studied.
The atlas of spectral simulations presented here will serve as a guide
to future reverberation campaigns.

\acknowledgments
Support for {\it HST} program number GO-13330 was provided by NASA through a grant from 
the Space Telescope Science Institute, which is operated by the Association of Universities
for Research in Astronomy, Inc., under NASA contract NAS5-26555. We thank NSF (1816537, 1910687), NASA (17-ATP17-0141, 19-ATP19-0188), and STScI (HST-AR-15018, HST-AR-14556). 
M.C. acknowledges support from STScI (HST-AR-14556.001-A), NSF (1910687), and NASA (19-ATP19-0188). MCB gratefully acknowledges support from the NSF through grant AST-2009230 to Georgia State University.
M.D.\ and G.F.\ and F. G.\  acknowledge support from the NSF (AST-1816537), NASA (ATP 17-0141),
and STScI (HST-AR-13914, HST-AR-15018), and the Huffaker Scholarship.
M.M. is supported by the Netherlands Organization for Scientific 
Research (NWO) through the Innovational Research Incentives Scheme Vidi grant 639.042.525.

 \end{document}